\documentclass{aa}
\usepackage{graphicx}
\usepackage{txfonts}
\usepackage{hyperref}
\usepackage{orcidlink}
\usepackage[normalem]{ulem}
\hypersetup{colorlinks = true, citecolor = {blue}, urlcolor= {blue}}
\def\gapprox{\;\rlap{\lower 3.0pt                       % approximately smaller
        \hbox{$\sim$}}\raise 2.5pt\hbox{$>$}\;}
\def\lapprox{\;\rlap{\lower 3.1pt                       % approximately smaller
        \hbox{$\sim$}}\raise 2.7pt\hbox{$<$}\;}

\newcommand{\be}{ \begin{equation} }
\newcommand{\ee}{\end{equation}}

\newcommand{\ben}{\begin{enumerate}}
\newcommand{\een}{\end{enumerate}}

\renewenvironment{thebibliography}[1]{\begin{oldthebibliography}{#1}\setlength{\parskip}{0ex}\setlength{\itemsep}{0ex}}{\end{oldthebibliography}}

\usepackage{diagbox}
\usepackage{siunitx}
\newcommand{\orcid}[1]{\href{https://orcid.org/#1}{\protect\includegraphics[width=8pt]{./figures/orcid.pdf}}}
\makeatletter
\renewcommand*\aa@pageof{, page \thepage{} of \pageref*{LastPage}}
\makeatother

\usepackage[dvipsnames]{xcolor}
\usepackage{booktabs}

\newcommand{\PGPU}{$\varphi\mathtt{-GPU}$}
\newcommand{\PGRAPE}{$\varphi\mathtt{-GRAPE}$}
\newcommand{\Msol}{\text{M}_\odot}

\begin{document}

    \title{Dynamics of spinning supermassive black hole binaries: I. \\ A direct $N$-body study}

    \author{
	    Navonil Saha\inst{1}\thanks{Fellow of the International Max Planck Research School for Astronomy and Cosmic Physics at the University of Heidelberg (IMPRS-HD)}%\orcidlink{0000-0002-7304-6865}
        \and
        Margarita Sobolenko\inst{2}%\orcidlink{0000-0003-0553-7301}
        \and
    	Peter Berczik\inst{2}%\orcidlink{0000-0003-4176-152X}
	    \and
        Andreas Just\inst{1}%\orcidlink{0000-0002-5144-9233}
        \and
        Walter Dehnen\inst{1}
	}

\institute{
{Zentrum f\"ur Astronomie der Universit\"at Heidelberg, Astronomisches Rechen-Institut, M\"{o}nchhofstr. 12-14, 69120, Heidelberg, Germany, \email{\href{mailto:navonil.saha@uni-heidelberg.de}{navonil.saha@uni-heidelberg.de}}}
\and
{Main Astronomical Observatory, National Academy of Sciences of Ukraine, 27 Akademika Zabolotnoho St, 03143 Kyiv, Ukraine } %\email{\href{mailto:berczik@mao.kiev.ua}{berczik@mao.kiev.ua}}
%\and
%{Nicolaus Copernicus Astronomical Centre, Polish Academy of %Sciences, ul. Bartycka 18, 00-716 Warsaw, Poland}
}

\date{Received 13 November 2026}

  \abstract
   {
    The dynamical evolution of supermassive black hole (SMBH) binaries in dense stellar environments may be influenced by the individual and mutual spin configurations of the black hole through spin–orbit and spin–spin couplings. We investigate the impact of spin on key dynamical properties, including binary hardening, orbital evolution, and energy exchange with the surrounding stellar population of an SMBH binary with masses \(m_1=3\cdot10^8\,\mathrm{M_\odot}\) and \(m_2=8\cdot10^7\,\mathrm{M_\odot}\), corresponding to a mass ratio \(q\simeq3.75\). We perform direct \(N\)-body simulations including post-Newtonian (PN) corrections up to 2.5PN order between the black hole-black hole interactions, considering seven spin configurations, including a non-spinning model, and examine the effects of numerical resolution on the energy transferred to the stars. For the adopted initial conditions, the binary merger times vary only by  $1.7\%$ of the $\sim3.8$~Myr total inspiral time. The hardening rates and cumulative energy transferred to the stellar population also show modest, non-monotonic variations across the spin configurations, with no clear systematic dependence on spin magnitude or orientation. The energy gained by the stars is consistent with the corresponding energy lost by the binary during the same encounters, including that radiated in gravitational waves, to within approximately $1\%$, which sets the level of the energy error of the balance. We further find that prograde stellar orbits account for roughly three quarters of the cumulative energy gain, a preference that is also present in the non-spinning model and therefore cannot be attributed to black-hole spin.
    Our results indicate that, for the binary parameters and stellar model considered here, spin has a limited influence on the global merger timescale and overall energy redistribution, while stellar interactions remain central to the hardening process.

   }


   \maketitle
%
%-------------------------------------------------------------------
%%%%%%%%%%%%%%%%%%%%%%%%%%%%%%%%%%%%%%%%%%%%%%%%%%%%%%%%%%%%%%%%%%%%%%%%
\section{Introduction}
It is now well known that most massive galaxies host in their center a supermassive black hole (SMBH) \citep{2013ARA&A..51..511K}. These SMBHs have masses in the range of $10^5$-$10^{10}\,\mathrm{M_\odot}$, which correlate strongly with the velocity dispersion of their host galaxy, implying that the host galaxies and their central SMBHs grow together and suggest a co-evolutionary connection between them such as the empirical $M_{\mathrm{BH}}$–$\sigma$ relation which links the SMBH mass $M_{\mathrm{BH}}$ to the stellar velocity dispersion $\sigma$ of the galactic bulge \citep{2000ApJ...539L...9F, 2000ApJ...539L..13G, 2013ApJ...764..184M}. When the galaxy grows, $\sigma$ increases and hence subsequently the $M_{\mathrm{BH}}$ also increases.

Various studies have been carried out using large-scale Hydrodynamical cosmological simulations to study the properties of black holes present inside the gravitationally bound galaxies such as the IllustrisTNG
\citep{2014Natur.509..177V,2018MNRAS.475..676S} or EAGLE \citep{2015MNRAS.446..521S} simulations.  Cosmological simulations are in general excellent for studying large-scale structure formation (e.g., galaxies and dark matter halos), galaxy evolution, and SMBH growth over the cosmic time of billions of years. However, they struggle when it comes to resolving the detailed dynamics of SMBH binaries \citep{2010A&ARv..18..279V, 2017MNRAS.464.3131K, 2017MNRAS.470.1121T}. They typically cover tens to hundreds of Mpc in size, meaning their smallest resolved scales are typically kpc or at best tens of parsecs in high resolution zoom-in runs, but this is insufficient to resolve the sub-parsec separations of SMBH binaries. Similarly,
the temporal resolution is too poor to track the rapid dynamical interactions \citep{2019MNRAS.483.3336T, 10.1093/mnrasl/slz091}.
In these simulations the SMBHs are usually repositioned to the minimum of the local potential well to avoid artificial wandering caused by limited resolution \citep{2015MNRAS.451.1868T}. The SMBHs are artificially seeded in halos, where their masses exceed a threshold of $\sim 10^5\,\Msol$ and their growth and feedback are primarily driven by gas accretion, modeled using semi-analytic subgrid prescriptions \citep{2005MNRAS.361..776S, 2009MNRAS.398...53B}, which often assume simple accretion and feedback models.
Such simulations cannot capture the full complexity of SMBH binary interactions, but instead apply simple recipes to merge the SMBHs when their separation falls below values at which the simulation can still resolve the dynamics \citep{2017MNRAS.471.4508K}. \citet{2016ApJ...828...73K} was the first to perform a full simulation starting with initial conditions drawn from a cosmological simulation of galaxy formation until the final merger of the SMBH binary by gravitational wave (GW) emission. 

When two galaxies merge, their central SMBHs eventually form a binary system \citep{1980Natur.287..307B, 2007Sci...316.1874M}. This process occurs in multiple stages. Firstly, as the two galaxies merge, their SMBHs, embedded in their respective stellar bulges, sink toward the center of the newly merged galaxy. This first stage is driven by dynamical friction \citep{1943ApJ....97..255C} against the background stars and dark matter, which is efficient at separations of 1-10\,kpc and helps in bringing the SMBHs close to the galactic nucleus by extracting energy from them \citep{1943ApJ....97..255C, 2001ApJ...563...34M}. Once the separation reaches $\sim10-100\,$pc, the two SMBHs form a gravitationally bound binary orbiting within a dense stellar background, where continued interactions with stars remove energy and angular momentum from the bound system \citep{1996NewA....1...35Q, 2010ApJ...719..851S}. 

The next phase is accompanied by the three-body interactions, or the hardening phase, where the binary hardens as stellar particles interact with it and are then ejected from the system via gravitational slingshot, shrinking the binary further as it loses angular momentum \citep{2004ApJ...606..788M, 2005LRR.....8....8M, 2006ApJ...642L..21B}. For spherical systems, stellar angular momenta are conserved and only stars with sufficiently small values come close to the SMBHs. Once this reservoir is depleted, no more slingshot interactions occur and the hardening of the SMBH binary stalls at $\sim1\,$pc, an effect known as the `final parsec problem' \citep{2003ApJ...596..860M, 2005ApJ...633..680B}. However, real systems deviate from spherical symmetry and the stellar angular momenta are not conserved, such that the reservoir of stars that eventually come close to the SMBHs is large and is not depleted \citep{2006ApJ...642L..21B,2011ApJ...732...89K,2011ApJ...732L..26P,2013ApJ...773..100K, 2015ApJ...810..139H, 2017ApJ...840...53R,2017MNRAS.464.2301G,2017MNRAS.464.3131K, 2018MNRAS.477.2310B}. 

Once the binary reaches separations of $\sim10^{-3}-10^{-2}$~pc, relativistic effects become important. GW emission extracts energy and angular momentum, shrinking the orbit and leading to coalescence within a timescale that depends on the binary’s mass ratio, eccentricity, and spin \citep{PhysRev.136.B1224,2004PhRvD..69h2005B,2004ApJ...611..623S}. In the final merger the SMBHs coalesce, thereby emitting a powerful burst of sub-Hz gravitational waves detectable by future space-based GW observatories like Laser Interferometer Space Antenna (LISA) and Pulsar Timing Arrays (PTAs) \citep{2017arXiv170200786A,2019MNRAS.486.4044B, 2023ApJ...951L...8A, 2023A&A...678A..50E, 2024arXiv240207571C}.

Due to the limitations of cosmological simulations in capturing the detailed dynamics of SMBHs down to sub-parsec scale, high-resolution gravitodynamical $N$-body simulations offer a more precise and alternative approach, to accurately model close binary interactions, including stellar scattering and dynamical friction \citep{2012ApJ...756...30K}. $N$-body simulations use a direct, particle-based approach for modeling gravitational interactions \citep[see][for a review]{2011EPJP..126...55D}, resolving the small scale dynamics of the SMBH binary with their surrounding stars on sub-parsec scales \citep{2012ASPC..453..223S, 2016IAUS..312...92V, 2021A&A...652A.134S, 2022A&A...665A..86B, 2026A&A...713A...5S}. In the final merger phase, when relativistic effects become important, these simulations usually include post-Newtonian (PN) corrections to study the merger dynamics of the binary and also model the eccentricity evolution, which impacts the merger timescales and ultimately results in GW emission \citep{2002PhRvD..65h4011G, PhysRevD.71.129902,Blanchet2024}. High resolution $N$-body re-simulations of the early formation phase of an elliptical galaxy with initial conditions drawn from massive galaxy mergers of cosmological simulations have already shown a very efficient hardening of the binary with coalescence within only $\sim10$~Myr after the galactic cores have merged \citep{2012ApJ...749..147K, 2015ApJ...798..103K, 2016ApJ...828...73K}.

After its mass, the only other relevant parameter of a black hole is its spin\footnote{According to the No-Hair Theorem \citep{2005PhRvD..72h4013H, 2012LRR....15....7C}, electric charge is another parameter, but completely unimportant since easily and naturally neutralized via accretion.}, which impacts the innermost stable circular orbit \citep[ISCO;][]{1972ApJ...178..347B}. This in turn affects the structure of any accretion disc, and hence also black hole growth and AGN feedback, as well as tidal disruption events \citep{1998A&A...331L...1S, 2005Natur.433..604D, 2006ApJS..163....1H, 2015ARA&A..53..115K}. The spin of SMBHs in a coalescing binary also affects the GW recoil, potentially ejecting the remnant black hole from the host galaxy \citep{2007ApJ...659L...5C, 2007PhRvL..98w1101G}.

Until now, there have been only a handful of studies of SMBH spins using either observational constraints \citep{2013CQGra..30x4004R, 2013mams.book.....B, 2017arXiv170307182W, 2021ARA&A..59..117R, 2021SSRv..217...65B, sisk2026} or large-scale cosmological and magnetohydrodynamical simulations \citep{2018MNRAS.477.3807F,10.1093/mnras/stz2836,2024A&A...685A..92S, 2024MNRAS.52710867B,2025MNRAS.536.1838B, 2025ApJ...980..136R} that lack an accurate model of the actual dynamics of these binaries, as discussed previously. Moreover, the spin magnitudes and orientations of SMBHs are themselves uncertain: they are expected to evolve through gas accretion and mergers over cosmic time \citep{2005ApJ...620...69V, 2008ApJ...684..822B}, and gas-rich mergers may favor alignment of the spins with the orbital angular momentum \citep{2007ApJ...661L.147B, 2010MNRAS.402..682D}. It is therefore still uncertain how spin changes the dynamical evolution and the final outcome of merging SMBH binaries.

In this work, we investigated the effect of spin on the dynamics of the SMBH binary using detailed $N$-body simulations. We adopted the same initial conditions as those used by \cite{2016ApJ...828...73K} and \cite{2021A&A...649A..41A}, and then analyzed if different spin configurations influence the dynamics and overall evolution of the binaries. Additionally, we explored the impact of varying the number of interacting stellar particles, using a direct-summation $N$-body code to model the evolution after the hardening phase until the final binary merger.

This paper is organized as follows. In Sect.~\ref{sec:Model setup}, we describe the details of the spin dynamics, our simulation methods and our initial conditions. In Sect.~\ref{sec:Results}, we present our results, while Sect.~\ref{sec:Conclusions and Discussions} discusses and summarizes our findings.
%%%%%%%%%%%%%%%%%%%%%%%%%%%%%%%%%%%%%%%%%%%%%%%%%%%%%%%%%%%%%%%%%%%%%%%

%%%%%%%%%%%%%%%%%%%%%%%%%%%%%%%%%%%%%%%%%%%%%%%%%%%%%%%%%%%%%%%%%%%%%%%
\section{Model setup} 
\label{sec:Model setup}

The details of the spin dynamics are presented in Sect.~\ref{Spin and Post-Newtonian approximation}, followed by a short description of our $N$-body methods in Sect.~\ref{Code description} and a discussion of our initial conditions in Sect.~\ref{Initial conditions}.

We study the effect of the spin configuration as well as of the numerical resolution (as parameterized by number of stellar particles) on the evolution of the binary, as quantified by dynamical parameters like the black-hole separation ($\Delta R_{\mathrm{BH}}$), semi-major axis (a), eccentricity ($e$), orbital period ($T$), cumulative energy change ($\Delta E_{*}$), specific energy change, stellar hardening rate ($s$), and the overall energy balance of the binary.

\subsection{Spin and post-newtonian approximation} \label{Spin and Post-Newtonian approximation}
In the following sections, we denote with $m_{1,2}$ and $\vec{r}_{1,2}$ the masses and positions of the black holes, while their separation and distance with $\vec{r}\equiv\vec{r}_1-\vec{r}_2$ and $r=|\vec{r}|$, respectively. The momentum conjugate to $\vec{r}$ is $\vec{p}=\mu\dot{\vec{r}}$, where $\mu\equiv m_1m_2/M$ is the reduced mass and $M=m_1+m_2$ is the total mass. We use a hat to denote a unit vector, i.e.\ $\hat{\vec{r}}\equiv\vec{r}/r$.

\subsubsection{Black-hole spin} \label{Black-hole spin}
The magnitude of the spin of a SMBH, i.e.\ its angular momentum with respect to its center of mass frame \citep{2011ApJ...728...12L, 2017Natur.548..426F}, is commonly expressed in terms of a dimensionless spin parameter $0\le a_S \le 1$ as: 
\begin{equation}
    |\vec{S}_{\mathrm{BH}}| = a_S \frac{G{m}^2}{c}
\end{equation}
where, $m$ is the mass of the SMBH, $G$ the gravitational constant, and $c$ the speed of light. 
This spin parameter measures how fast a black hole is rotating, where, $a_S=0$ corresponds to a non-rotating (Schwarzschild) black hole, $a_S=1$ for an extremal Kerr black hole with a maximum possible spin, while $a_S > 1$ is unphysical as it corresponds to a naked singularity (violating the cosmic censorship conjecture).

The orientations of the black-hole spins with respect to the angular momentum $\vec{L}\equiv\vec{r}\times\vec{p}$ of their mutual orbit are also relevant. These are quantified by the `projected spin parameters' \citep{2017Natur.548..426F} of the individual black holes,
\begin{equation}
    \label{eq:chi1,2}
    \chi_{1,2} = a_{S_{1,2}}\cos\theta_{1,2}
    = \frac{c}{Gm^2_{1,2}}\,\vec{S}_{1,2}\cdot\hat{\vec{L}},
\end{equation}
where $\vec{S}_{1,2}$ are the black hole spins and $\theta_{1,2}$ the angles between the $\vec{S}_{1,2}$ and $\vec{L}$. Then the combined `effective spin parameter' \citep{2011PhRvL.106x1101A, 2016PhRvL.116x1102A, 2017Natur.548..426F, 2020A&A...635A..97B} of the SMBH binary is :
\begin{equation}
    \chi_{\mathrm{eff}}
    \equiv \frac{m_{1}\chi_1 +m_{2}\chi_2}{M}
    =\frac{c}{GM}\left(\frac{\vec{S}_{1}}{m_{1}}+\frac{\vec{S}_{2}}{m_{2}}\right)\cdot\hat{\vec{L}}
\end{equation}
Since the dimensionless spin parameters are bounded between $0\le a_{S_{1,2}}\le1$, the projected and effective spin parameters are bounded too by $-1 < \chi_{1},\,\chi_{2},\,\chi_{\mathrm{eff}} < 1$. $\chi_{\mathrm{eff}}>0$, if the spins are mostly aligned with the orbital angular momentum. Such alignment also 
leads to higher final spins of the remnant with a high positive $\chi_{\mathrm{eff}}$ \citep{2001PhRvD..64l4013D, 2008PhRvD..78d4021R, 2009ApJ...704L..40B}. If the individual black hole spins are randomly oriented with respect to the orbital angular momentum, then $\chi_{\mathrm{eff}}\sim 0$, and the system behaves more like a binary of non-spinning black holes \citep{2018PhRvD..98h4036G, 2019ApJ...882L..24A} showing precession for misaligned spins \citep{2014PhRvL.113o1101H}. Finally, $\chi_{\mathrm{eff}} < 0$ for spins that are anti-aligned with the orbital angular momentum. 

%%%%%%%%%%%%%%%%%%%%%%%%%%%%%%%%%%%%%%%%%%%%%%%%%%%%

%%%%%%%%%%%%%%%%%%%%%%%%%%%%%%%%%%%%%%%%%%%%%%%%%%%

\subsubsection{Post-newtonian approximation}
\label{Post-Newtonian approximation}
The PN formalism expands the equations of motion of a compact binary in powers of $1/c$ to incorporate the relativistic corrections \citep[e.g.][]{Blanchet2024}.
In PN theory, the Hamiltonian of a binary system in its barycentric frame \footnote{we consider, \vec{c}=1 in the expansions for compactness} 
is given as follows \citep{2000AnP...512..378J, 2001PhRvD..64l4013D}
\begin{align}
    H=H_{\mathrm{N}} + \frac{H_{1\text{PN}}}{c^2} + \frac{H_{\text{SO}}}{c^3} + \frac{H_{2\text{PN}}}{c^4}  + \frac{H_{\text{SS}}}{c^4}  + \frac{H_{2.5\text{PN}}}{c^5}  + \dots
\end{align}
Here, the Newtonian Hamiltonian is,
\begin{equation}
    H_{\text{N}} = \frac{p^2}{2\mu}-\frac{GM\mu}{r}
\end{equation}
where, $p\equiv|\vec{p}|$. 
The first post-Newtonian correction
accounts, among other effects, for periapsis advance and is given by \citep{1999PhRvD..59h4006B}, 
\begin{align}
    H_{\text{1PN}}=\frac18(3\nu-1)\frac{p^4}{\mu^3} - \frac{GM\mu}{2r}\left[(3+\nu)\frac{p^2}{\mu^2}+\nu\frac{(\hat{\vec{r}}\cdot\vec{p})^2}{\mu^2}\right] + \frac{G^2\!M^2\mu}{2r^2},
\end{align}
where, $\nu\equiv\mu/M$ is the symmetric mass ratio. The second PN correction \citep{1999PhRvD..59h4006B, 2000AnP...512..378J}
\begin{align}
    H_{2\mathrm{PN}} &= \frac{1}{16} \left(1 - 5 \nu + 5 \nu^2 \right) \frac{p^6}{\mu^5} \nonumber \\
    &+ \frac{GM}{8 r} \left[\left(5 - 20\nu - 3\nu^2\right) \frac{p^4}{\mu^3} - 2 \nu^2 \frac{p^2 (\hat{\vec{r}}\cdot \vec{p})^2}{\mu^3} - 3 \nu^2 \frac{(\hat{\vec{r}}\cdot \vec{p})^4}{\mu^3} \right] \nonumber \\ 
    &+\frac{G^2\!M^2}{2 r^2} \left[ (5 + 8 \nu) \frac{p^2}{\mu} + 3 \nu \frac{(\hat{\vec{r}}\cdot \vec{p})^2}{\mu} \right] - \frac{G^3\!M^3\mu}{4 r^3} (1 + 3 \nu),
\end{align}
further modifies the periapsis precession.

The leading spin-dependent term, the spin–orbit coupling Hamiltonian, which appears at 1.5PN order ($\propto c^{-3}$),
\begin{equation}
    H_{\text{SO}}=\frac{G}{r^3}\left[\left(2+\frac{3}{2}\frac{m_2}{m_1}\right)\vec{S_{1}}+\left(2+\frac{3}{2}\frac{m_1}{m_2}\right)\vec{S_{2}}\right]\cdot(\vec{r}\times \vec{p}),
\end{equation}

\citep{1979GReGr..11..149B, 1995PhRvD..52..821K, 2001PhRvD..63d4006T, 2006PhRvD..74j4033F, 2018PhRvD..97f4010V},
describes the interaction between the black-hole spins and the orbital angular momentum. This coupling arises because the dragging of spacetime by a spinning black hole perturbs the motion of its companion, driving precession of the orbital plane \citep{1994PhRvD..49.6274A}. 

The spin–spin coupling, which enters at 2PN order ($\propto c^{-4}$), is given by,
\begin{equation}
\begin{aligned}
H_{\text{SS}} ={}&
\frac{G}{r^3}
\left[
3\left(\vec{S_1}\cdot\hat{\vec{r}}\right)
 \left(\vec{S}_2\cdot\hat{\vec{r}}\right)
-\vec{S}_1\cdot\vec{S}_2
\right]+
%\\
%&+
\frac{G}{2 r^3}\frac{m_2}{m_1}
\left[
3\left(\vec{S}_1\cdot\hat{\vec{r}}\right)^2
-\vec{S}_1^2
\right]
\\
&+
\frac{G}{2 r^3}\frac{m_1}{m_2}
\left[
3\left(\vec{S}_2\cdot\hat{\vec{r}}\right)^2
-\vec{S}_2^2
\right],
\end{aligned}
\end{equation}

\citep{1975PhRvD..12..329B,1979GReGr..11..149B, 1993PhRvD..47.4183K, 1995PhRvD..52..821K, 2001PhRvD..64l4013D, 2006PhRvD..74j4005B}, and induces spin-dependent orbital variations that can either drive the growth or damping of the binary eccentricity \citep{2004PhRvD..70l4020S}.

Finally, the 2.5PN dissipative term,

\begin{align}
        H_{2.5\text{PN}} = \frac{2G}{5}\dddot{I}_{ij}\left(\frac{p_{1i}p_{1j}}{m_1}+\frac{p_{2i} p_{2j}}{m_2}-\frac{G m_1\ m_2}{r_{12}}\hat{r^i_{12}}\hat{r^j_{12}}\right),
\end{align}
where $I_{ij}$ is the Newtonian mass quadrupole moment of the binary,
    $I_{ij} =
    \sum_{a=1}^{2}
    m_a
    \left(
    x_a^i x_a^j
    -\frac{1}{3}\delta_{ij}x_a^2
    \right)$,
and ($\dddot{I}_{ij}$) denotes its third time derivative \citep{2018LRR....21....7S}, it introduces the leading-order dissipative effects due to GW emission. This term drives the inspiral of the binary by extracting orbital energy and angular momentum through gravitational radiation \citep{1995PhRvD..51.5360B, Blanchet2024}.

%%%%%%%%%%%%%%%%%%%%%%%%%%%%%%%%%%%%

\subsection{Code description} 
\label{Code description}

For the global dynamical integration we used the direct $N$-body code \PGPU\footnote{$N$-body code \PGPU:\\~\url{https://github.com/berczik/phi-GPU-mole}} \citep{Berczik2011, BSW2013, Berczik2024ascl}. The code solves the equations of motion of all particles with direct summation of the pairwise forces, using a Hermite integration scheme with hierarchical individual block time steps \citep{1992PASJ...44..141M, 2003gnbs.book.....A}. Fourth-, sixth- and eighth-order Hermite schemes are available \citep{Nitadori2008, Berczik2011}; in this work we use the fourth-order scheme. \PGPU\ is written in C++ and is based on the earlier C code \PGRAPE, originally developed for GRAPE6a clusters \citep{2007NewA...12..357H}; the integration of the particle orbits and the time-step criterion follow the serial high order CPU $N$-body code \citep{Nitadori2008}. The version of the code with a special treatment of the SMBH particles has been thoroughly tested and used for a wide range of models with multiple SMBHs \citep{Sedda2019, 2021A&A...649A..41A, Koehn2023A&A, 2022A&A...665A..86B, Berczik2024A&A, 2025A&A...703A..23B}; for the present project it has been further adapted to the computational requirements of the spin runs.

The forces are computed on GPUs through the NVIDIA CUDA library, and the multi-GPU parallelization is done with MPI (Message Passing Interface). Each MPI process drives one GPU, and several MPI processes per node are used, so that both the multi-core CPUs and all GPUs of a node are employed. The particles are distributed evenly over the MPI processes; in every block step each process computes on its GPU the partial forces exerted by its own particles on the active ($i$-) particles, and the full forces on the active particles are obtained with a global MPI reduction.

The individual time steps are computed with the generalized Aarseth-type criterion proposed by \citet{Nitadori2008}. In this work we use a value of $\eta_{*} = 0.15$ for the timestep accuracy parameter for the background particles. For the SMBH particles we introduce a much smaller value of $\eta_{\rm BH} = 0.02$.

The time steps of the SMBH particles are, in addition, minimised and forcibly synchronised: after every block step each SMBH particle receives the smallest time step of all particles of that block. In this way all SMBH particles are integrated together in every block step, without any additional block steps, and their time step is never larger than that of any other particle.

The effective Plummer type gravitational softening of the pair: $ \varepsilon_{ij}^2 = \alpha\ \cdot 0.5 \cdot (\varepsilon^{2}_{i} + \varepsilon^{2}_{j}) $, where $\varepsilon_{i}$ is the individual softening of the particle type and $\alpha$ is a reduction factor: $\alpha = 1$ for all star--star interactions and $\alpha < 1$ for the interactions between stars and SMBHs, which allows the dynamical evolution of the SMBHs to be followed accurately up to the PN merging phase. The SMBH particles have exactly zero softening, $\varepsilon_{\rm BH} = 0$, so that the SMBH--SMBH interactions are computed without any softening.

In this work we use $\varepsilon_{*} = 10^{-1}$ pc and $\alpha = 10^{-4}$ for the star--SMBH interactions, which gives effective softening lengths for this mixed gravitational interaction of $\sim7 \cdot 10^{-4}$ pc. Our approach is similar to other recent $N$-body implementations with PN terms \citep[see][]{Rantala2018, Rantala2019, Mannerkoski2019}.

The PN corrections are included as additional acceleration terms for the interactions between the SMBHs only; the star--SMBH interactions are treated in pure Newtonian dynamics. This is a simplifying assumption, and we leave the relativistic effects of star--SMBH interactions to future work. The code includes the PN terms up to order 3.5PN, including the spin--orbit and spin--spin interactions \citep{2013MNRAS.434.2999B, Sobolenko2017, 2021A&A...652A.134S, Sobolenko2022}; the acceleration terms can be derived from the corresponding PN Hamiltonians \citep[see also][]{Blanchet2024}. Among them, the 2.5PN term, i.e. of order $(v/c)^5$, is the leading dissipative term, which carries away most of the orbital energy of the binary SMBH through the emission of gravitational waves \citep{2018A&A...615A..71K, Khan2018ApJ, Khan2026A&A, Berczik2024A&A}. The simulation is stopped when the separation of two SMBHs falls below four times the sum of their Schwarzschild radii, $4\,(R_{{\rm S},1} + R_{{\rm S},2})$ with $R_{\rm S} = 2GM/c^2$; this time is taken as the merging time of the binary SMBH.

%%%%%%%%%%%%%%%%%%%%%%%%%%%%%%%%%%%%

%%%%%%%%%%%%%%%%%%%%%%%%%%%%%%%%%%%%%%%%%%%%%%%%%%%%%%%%%%%%%%%%%%%%

%%%%%%%%%%%%%%%%%%%%%%%%%%%%%%%%%%%%%%%%%%%%%%%%%%%%%%%%%%%%%%%%%%%%
\subsection{Initial conditions} \label{Initial conditions}
We initialize our system based on the previous $N$-body runs in \cite{2016ApJ...828...73K} and later updated by \cite{2021A&A...649A..41A}, to simulate the late merging phase of two SMBHs with masses,  $m_{1} = 3 \cdot 10^{8} \mathrm{M_\odot}$ and $m_{2} = 8\cdot 10^{7} \mathrm{M_\odot}$, and corresponding mass ratio $q=m_1/m_2=3.75$. The binary is embedded in a dense stellar environment modeled with $N$ particles. 

From the previous Argo cosmological hydrodynamical galaxy merger simulation \citep{2015MNRAS.446.1939F, 2015MNRAS.446.1957F}, \cite{2016ApJ...828...73K} extracted a spherical region of $\sim5$ kpc at time $t_0=21.5$~Myr in the merger phase, by a particle splitting technique to enhance the resolution around the more massive SMBH when the separation of the two SMBHs reaches about 300~pc. This generated an initial model with $6\cdot10^6$~particles. During the following $\approx14$~Myr of evolution, the inner region contracted, as the SMBHs sink from $\approx0.2$\,kpc to a bound binary at $\sim10^{-2}$\,pc from the centre, while the outer parts expanded, so that the halo particles are extended to $\sim10$\,kpc.

For the final merger phase, we started our spin simulations from $t=35.45$~Myr with a binary separation of $\sim 450R_{\mathrm{s}}$. The initial semi-major axis and the eccentricity of the SMBH binary at that time are, $a\approx0.018\,$pc and $e\approx0.09$, respectively.

To determine the number of stellar particles required to robustly follow the SMBHs through their formation of a hard binary and merger, we performed a resolution study starting from $\sim 6\cdot10^6$~stellar particles and then gradually reduced the number of particles to $\sim 2.19\cdot10^5$ for a particular spin configuration. We find that $\sim 1.5\cdot10^6$~particles provide sufficient resolution to capture the binary dynamics while keeping the computational cost manageable, since with our direct-summation approach, several simulations of different SMBH spin configurations are numerically too expensive for $6\cdot10^6$~ particles. Therefore we adopt $\sim 1.5\cdot10^6$ particles for all spin runs (see Sect.~\ref{Size of the stellar system}). All spin configurations are listed in Table~\ref{table1}. Runs $S_{\mathrm{E1}}$-$S_{\mathrm{E5}}$ adopt near-maximal spin magnitudes for both black holes, with different mutual orientations. These high spin magnitudes are chosen to maximize the potential influence of spin effects on the binary dynamics. Run $S_{\mathrm{E6}}$ has half the spin magnitude of $S_{\mathrm{E1}}$ for both black holes, while  $S_{\mathrm{0}}$ is non-spinning. Runs $S_{\mathrm{A1}}$-$S_{\mathrm{H1}}$ use the $S_{\mathrm{E1}}$ spin orientation but different numbers of stellar particles and are used for the resolution study. As the particle data did not contain any spin information, so we did not change anything to the main data file and kept everything constant according to the rotating system apart from just cutting the particles in steps from the tail of the main file. %We also adopted the global gravitational softening of $\epsilon\sim10^{-4}$\,pc for general star-black hole interactions.

\begin{figure}[]
   \centering
   \includegraphics[width=\linewidth]{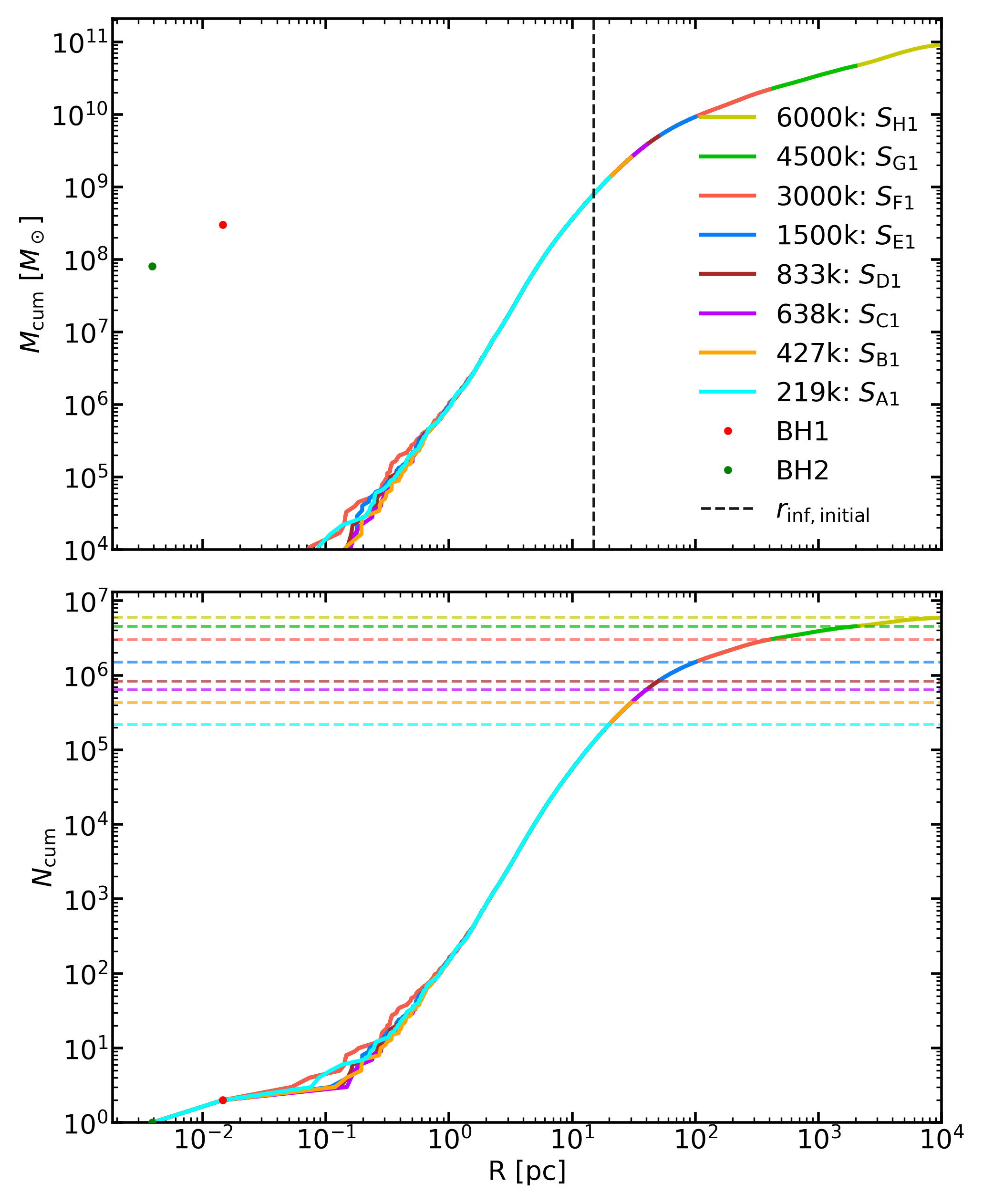}
   \caption{Cumulative mass and number distribution of the stellar system surrounding the SMBH binary for different number of interacting stellar particles. The initial influence radius $r_{\text{inf,initial}}$ is marked in vertical black dashed line. The total particle numbers are marked as dashed lines on the $N_\text{cum}$ axis. The two BHs are marked with red and green dots. Positions are given relative to the binary center of mass (see Sect.~\ref{Initial conditions}). 
   }
    \label{Cum-mass-num-dist}
\end{figure}
The influence radius, $r_{\mathrm{inf}}$ of a  SMBH binary is empirically given by the radius enclosing a stellar mass equal to twice the total mass of the binary as follows \citep{2004cbhg.symp..263M} :  
\begin{equation} \label{eq:rinf}
   M_* (r<r_{\mathrm{inf}})=2(m_1+m_2)
\end{equation}
Following the above relation, we plotted the cumulative mass and cumulative number profiles of the binary system with the SMBHs marked with two dots (see Fig.~\ref{Cum-mass-num-dist}). At $t=35.45$\,Myr, all particles are contained within an inner radius of $\sim10$~kpc, well beyond twice the influence radius of the binary, ensuring that the stellar environment relevant to binary hardening is fully sampled. All particles are centered on the center of mass (COM) of the binary system.

Before constructing the radial particle cuts, we computed the center of mass of the bound SMBH binary as a whole and recentered the particle sub-system containing the SMBHs and the surrounding stellar particles on this binary COM. This corrected binary-COM position was then adopted as the reference point for the initial conditions of the spin simulations.

The binary's orbital angular-momentum vector initially pointed in an arbitrary direction, making it inconvenient to specify the spin orientations in the original coordinate system. We therefore rotated the system so that the orbital angular-momentum vector lay along the z-axis. Then we took 7 different mutual spin configurations for different numbers of particles and rotated their vectors from the $z$-coordinate system to the original coordinate system. We used the new spin vector values as initial spins for SMBHs as given in Table~\ref{table1}.

\begin{table}[]
\caption{Different mutual spin configurations for the two SMBHs with respect to the binary orbit.} 
\begin{center}
\resizebox{\columnwidth}{!}{
\begin{tabular}{c|c|c|c|c} 
 \hline
 ID & $N$ & $r_\mathrm{cut}$ & $(c/Gm_1^2)\vec{S}_1$ & $(c/Gm_2^2)\vec{S}_2$ \\ [0.2ex] 
 \hline\hline 
 $S_{\mathrm{A1}}$ & 219k & 0.02 & $[+0.1,+0.1,+0.8]$ & $[-0.1,-0.1,-0.8]$ \\ 
 %\hline
 $S_{\mathrm{B1}}$ & 427k & 0.03 &$[+0.1,+0.1,+0.8]$ & $[-0.1,-0.1,-0.8]$ \\ 
 %\hline
 $S_{\mathrm{C1}}$ & 638k & 0.04 &$[+0.1,+0.1,+0.8]$ & $[-0.1,-0.1,-0.8]$ \\ 
 %\hline
 $S_{\mathrm{D1}}$ & 833k & 0.05 &$[+0.1,+0.1,+0.8]$ & $[-0.1,-0.1,-0.8]$ \\ 
 %\hline
 $S_{\mathrm{E1}}$ & 1500k & 0.1 &$[+0.1,+0.1,+0.8]$ & $[-0.1,-0.1,-0.8]$ \\ 
 %\hline
 $S_{\mathrm{F1}}$ & 3000k & 0.4 & $[+0.1,+0.1,+0.8]$ & $[-0.1,-0.1,-0.8]$ \\ 
 %\hline
 $S_{\mathrm{G1}}$ & 4500k & 2 & $[+0.1,+0.1,+0.8]$ & $[-0.1,-0.1,-0.8]$ \\ 
 %\hline
 $S_{\mathrm{H1}}$ & 6000k & 10 & $[+0.1,+0.1,+0.8]$ & $[-0.1,-0.1,-0.8]$ \\ 
 %\hline 
 $S_{\mathrm{E2}}$ & 1500k & 0.1 &$[+0.0,+0.9,+0.0]$ & $[-0.9,+0.0,+0.0]$ \\ 
 %\hline
 $S_{\mathrm{E3}}$ & 1500k & 0.1 &$[+0.9,+0.0,+0.0]$ & $[-0.9,+0.0,+0.0]$ \\ 
 %\hline 
 $S_{\mathrm{E4}}$ & 1500k & 0.1 &$[+0.0,+0.9,+0.0]$ & $[+0.0,-0.9,+0.0]$ \\ 
 %\hline  
 $S_{\mathrm{E5}}$ & 1500k & 0.1 &$[+0.0,+0.0,+0.9]$ & $[+0.0,+0.9,+0.0]$ \\ 
 %\hline  
 $S_{\mathrm{{E6}}}$ & 1500k & 0.1 &$[+0.05,+0.05,+0.4]$ & $[-0.05,-0.05,-0.4]$ \\ 
 %\hline 
 $S_\mathrm{0}$ & 1500k & 0.1 &$[+0.0,+0.0,+0.0]$ & $[+0.0,+0.0,+0.0]$ \\ 
 \hline
\end{tabular}}
\tablefoot{Here "+" indicates an aligned-spin and "$-$" indicates an anti-aligned spin with respect to spin vector of the orbit. The $r_\mathrm{cut}$ is given in kpc.}
\label{table1}
\end{center}
\end{table}
%%%%%%%%%%%%%%%%%%%%%%%%%%%%%%%%%%%%%%%%%%%%%%%%%%%%%%%%%%%%%%%

%%%%%%%%%%%%%%%%%%%%%%%%%%%%%%%%%%%%%%%%%%%%%%%%%%%%%%%%%%%%%%%
\section{Results}\label{sec:Results}
\subsection{Dynamical evolution of spinning SMBH binary}
\begin{figure*}[h]
   \centering
      \includegraphics[width=\textwidth]{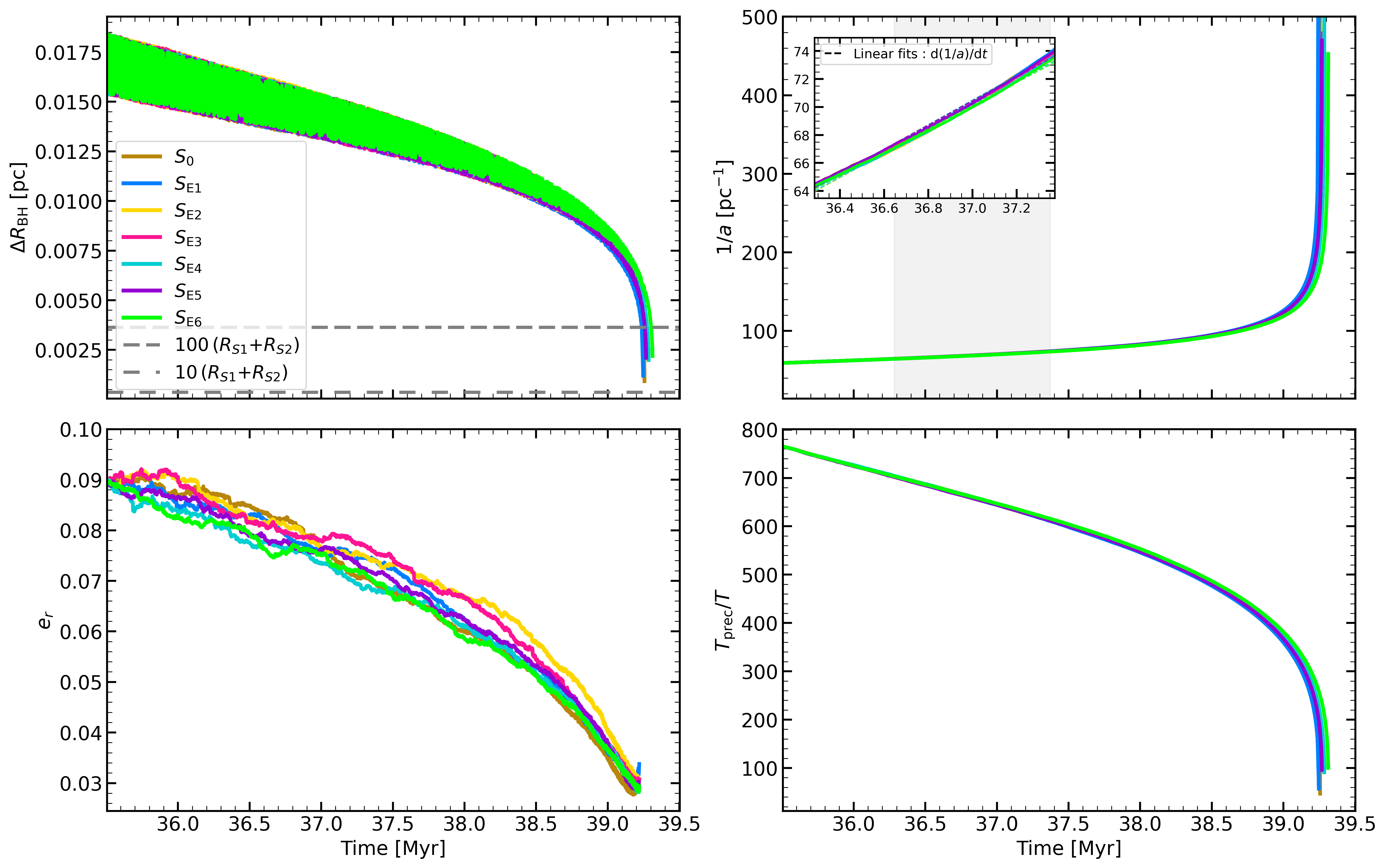}
   \caption{Time evolution of different orbital parameters. Top left: separation between the binary ($\Delta R_{\text{BH}}$).  Top right: Inverse of semi-major axis $(1/a)$. The shaded grey region indicates the time interval over which the hardening rate \(s\) is calculated. Bottom left: radial eccentricity of the binary ($e_r$). Bottom right: ratio of the precession timescale to the orbital period ($T_{\text{prec}}/{T}$).}
    \label{plots-1500k}%
\end{figure*}
In our work, we track the dynamics of the SMBH binary once their separation falls below the sub-parsec scale, comparing how different mutual spin configurations influence their evolution and merger time scales relative to the non-spinning case (see Fig.~\ref{plots-1500k}). We started our simulations with  $1.5\cdot10^6$ particles at an initial time of $t\sim35.45$~Myr and modelled the final merger phase of the SMBH binary until it reached a separation of a few Schwarzschild radii. On the sub-parsec scale in the final phase, the separation of the SMBHs decreases gradually, due to energy extraction via stellar slingshot interactions with the hard SMBH binary, ultimately hardening the binary towards coalescence. GW also begins to efficiently extract energy and angular momentum, for our binary, when the  separation reaches $\leq0.01$~pc, and the binary then tightens on a timescale of the order of a few million years until coalescence.

For our SMBHs, binary tightening begins at separation of around 0.018~pc and continues to harden for approximately 3.8~Myr from the initial time of 35.45~Myr to merger (see Fig.~\ref{plots-1500k}). As the binary tightens, its semi-major axis $(a)$ decreases according to the Keplerian orbits, and consequently the inverse semi-major axis $(1/a)$ increases. The evolution of $(1/a)$ therefore provides a direct measure of the stellar hardening of the binary. We quantify this hardening through the hardening rate $s \equiv \frac{d(1/a)}{dt}$, where $s$ measures the rate at which the inverse semi-major axis increases over time. In the regime where stellar interactions dominate the orbital evolution, the evolution of $(1/a)$ can be approximately described by a linear trend, with the slope of the $(1/a)$ curve providing an estimate of $s$. The measured values of $(s)$ for the different models are given in Table~\ref{table2}. 
The increase of $(1/a)$ reflects the primary loss of orbital energy through three-body interactions with the surrounding stars, with a small but non-negligible contribution from GW energy loss (included via 2.5PN corrections) as the binary tightens.

The differences in hardening rates between the spin configurations therefore provide a quantitative measure of the effect of black-hole spins on stellar hardening. A larger $(s)$ corresponds to a faster increase of $(1/a)$, and hence to a more rapid shrinkage of the binary and an earlier merger.

The exact merger timescale for all the configurations of the SMBH binaries are given in Table~\ref{table2}. On comparing the spinning and the non-spinning ($S_{\mathrm{0}}$) runs, we see that all models reach coalescence within a narrow time window. The full-spin model $S_{\mathrm{E1}}$ merges earliest at 39.247~Myr, while the half-spin model $S_{\mathrm{E6}}$ merges latest at 39.312~Myr. The non-spinning model $S_{\mathrm{0}}$ merges at 39.256~Myr and is therefore not the slowest case. The spread in merger times is only about $\Delta t_\text{merge}=39.312-39.247=0.065$~Myr, corresponding to a spread of only 0.17\% relative to the absolute simulation time. Thus, the present simulations reveal a small configuration dependence but do not support a strong monotonic relation between spin magnitude or spin orientation and the total merger time. The $S_{\mathrm{0}}$ model is almost indistinguishable in merger time from all other spin models. This indicates that the absence of spin-orbit coupling does not, for this particular initial condition, lead to a substantial delay in coalescence. 

\begin{table}[]
\caption{The actual merging times of the SMBH binary for the different spinning and non-spinning mutual configurations.}
\begin{center}
\begin{tabular}{c|c|c} 
 \hline
 {ID} & {$s$} & {$t_\text{merge}$} \\ [0.2ex] 
 \hline\hline
 $S_{\mathrm{E1}}$ & 8.783 & 39.247 \\ 
 %\hline 
 $S_{\mathrm{E2}}$ & 8.635 & 39.277 \\ 
 %\hline
 $S_{\mathrm{E3}}$ & 8.651 & 39.273 \\ 
 %\hline 
 $S_{\mathrm{E4}}$ & 8.560 & 39.285 \\ 
 %\hline  
 $S_{\mathrm{E5}}$ & 8.618 & 39.268 \\ 
 %\hline  
 $S_{\mathrm{E6}}$ & 8.332 & 39.312 \\ 
 %\hline 
 $S_{0}$ & 8.734 & 39.256 \\ 
 \hline
\end{tabular}
\tablefoot{The hardening rate $s$ is given in $\text{pc}^{-1}\; \text{Myr}^{-1}$and was calculated over 36.28-37.37~Myr; $t_\text{merge}$ is given in Myr.}
\label{table2}
\end{center}
\end{table}

As the binaries move towards the merger phase, the orbits become less and less eccentric, with \(e_r\) decreasing from $\approx$ 0.09 to $\approx$ 0.03 over the simulated time (Fig.~\ref{plots-1500k}). The eccentricity of a hard binary can be changed by stellar encounters \citep{1992MNRAS.259..115M, 1996NewA....1...35Q}. Here, the radial eccentricity \(e_r\) is defined according to the PN generalized quasi-Keplerian parametrization of \cite{2004PhRvD..70j4011M}. The orbital period of the binary is the time it takes to complete one full orbit around their common COM. As the binary tightens and hardens or its semi-major axis decreases, the time period also shortens and gradually decreases due to the $a^{3/2}$ dependence in the scale of a few years following $T=2\pi\sqrt{{a^3}/{GM}}$ \citep{Binney2008}. 
Following the orbital time, as the binary inspirals and the orbital separation decreases, relativistic spin couplings cause the orientation of the system to evolve through precession. The precession timescale characterizes the timescale over which the orbital angular momentum changes its orientation due to spin-orbit and spin-spin couplings \citep{2015PhRvD..92f4016G}. At leading PN order, the spin-orbit coupling dominates over the spin-spin coupling \citep{1995PhRvD..52..821K}. For misaligned spins, the orbital angular momentum $\vec{L}$ generally precesses about the total angular momentum $\vec{J}$, while the individual spins $\vec{S}_{1,2}$ undergo coupled precessional motion. In the generic case, the precession of $\vec{L}$ can also be accompanied by nutation \citep{PhysRevD.103.124026}. The precession therefore governs how quickly the orientation of the binary system evolves. It is related to the orbital period through the following equation, derived from \cite{PhysRevD.103.124026} for our models:
\begin{equation} \label{T_prec}
    T_{\mathrm{prec}}=\frac{T}{(1+\frac34q)\nu}\frac{a}{R_{\mathrm{s}}}
\end{equation}
where, $R_{\mathrm{s}} = 2GM/c^2$. For our mass ratio ($q=3.75$, $\nu\simeq0.166$), Eq.~(\ref{T_prec}) reduces to $T_{\rm prec}/T\simeq1.6\,a/R_{\rm s}$, so this ratio follows the shrinking semi-major axis and is independent of the spin magnitudes and orientations. The sub-plot in the bottom-right panel of Fig.~\ref{plots-1500k}, was obtained by evaluating this relation with the simulated $a$, and they are practically identical for all runs.

From this panel, $T_{\rm prec}/T$ starts at $\approx750$ at $t=35.45$~Myr and decreases monotonically as the binary hardens, reaching $\approx30$--$50$ just before the final plunge. Hence $T\ll T_{\rm prec}$
throughout the simulated merger phase, consistent with the PN approximation timescale hierarchy of \cite{PhysRevD.103.124026}. The binary completes between $\sim30$ and $\sim750$ orbits per precession cycle. The precession of the orbital plane and of the spins is thus much slower and hence more longer than the orbital motion, and the spin--orbit (1.5PN) and spin--spin (2PN) couplings act as a weak perturbation, with a very small precession angle per orbit, accumulating only slowly over many cycles. This agrees with the merger times and hardening rates discussed above, which show no systematic dependence on the spin configuration.
%%%%%%%%%%%%%%%%%%%%%%%%%%%%%%%%%%%%%%%%%%%%%%%%%%%%%%%%%%%%%%%%%%

%%%%%%%%%%%%%%%%%%%%%%%%%%%%%%%%%%%%%%%%%%%%%%%%%%%%%%%%%%%%%%%%%%%
\subsection{Energy balance of the binaries} \label{Energy Balance of the binaries}

The energy evolution of the SMBH binary embedded in a dense stellar environment is a fundamental diagnostic of its dynamical evolution toward coalescence. In our direct \(N\)-body simulations, stars interact gravitationally with the binary and continuously exchange energy and angular momentum with it. In particular, repeated three-body encounters can transfer orbital energy from the binary to individual stellar particles, some of which subsequently remain in quasi-bound orbits while others acquire sufficient energy to populate the high-energy tail and eventually escape. The cumulative energy transfer therefore provides a direct measure of the contribution of stellar interactions to the hardening of the SMBH binary.

To track this process quantitatively, we examine the overall energy balance of the SMBH binary along with their corresponding stellar particles. The total energy exchange of the stellar particle is computed by summing the changes in mechanical energy (kinetic + potential) of the stellar particles over all encounters with the binary, as follows:
\begin{equation}\label{Etot}
   \Delta E_{*(\mathrm{tot})}(t)= \sum_\mathrm{encounters} \left[E_{*}(t_{\mathrm{out}})-E_*(t_{\mathrm{in}})\right]
\end{equation}

Here, \(t_{\rm in}\) and \(t_{\rm out}\) denote the times at which a stellar particle enters and leaves the binary's encounter region, defined by a sphere of radius 10\,$a$. This summation includes all completed passages through the encounter region prior to the end of the simulation, meaning stars on quasi-bound orbits that have not yet escaped are fully accounted for in both the energy and inclination distributions. The quantity therefore measures the net energy deposited into the stellar component through the sequence of binary-star interactions and can be compared directly with the corresponding loss of orbital binding energy of the SMBH binary.

For each stellar particle, the specific mechanical energy is defined as: 
\begin{equation}
    E_*/m_*=\tfrac{1}{2}\varv^2+\Phi,
\end{equation}
where, $\varv$ is the velocity of the stellar particle relative to the binary, and $\Phi$ is the gravitational potential experienced by the particle. The specific energy is useful for characterizing the orbital state of a particle and for distinguishing whether a particle is bound $(E_*/m_*<0)$ or unbound $(E_*/m_*>0)$. In the following, $\Delta E_*/m_*$ denotes the change in specific energy of a star during one encounter. It is positive if the star gains energy from
the binary and negative if it loses energy to it.

\begin{figure}[]
    \includegraphics[width=\linewidth]{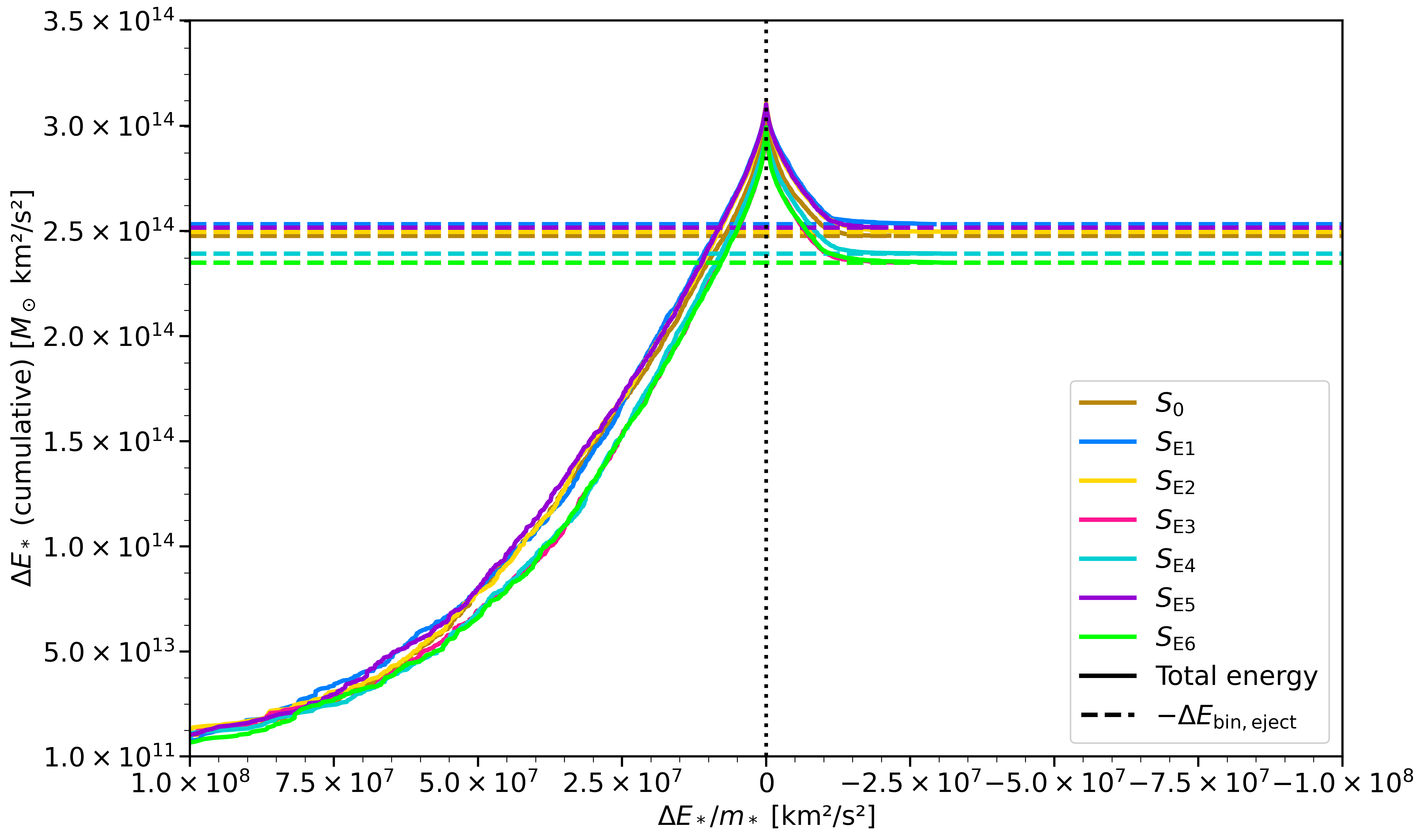}
    \caption{The cumulative energy change of the stellar particles as a function of their specific energy change showing the total energy balance for the different spin configurations. Please note that the x-axis is reversed.}
    \label{fig:energybalance}
\end{figure}

To verify that we identify the encounters responsible for the hardening, we compare the cumulative energy change of the stars with the total energy lost to the ejected stars from the binary system over the same period. Fig.~\ref{fig:energybalance} shows the cumulative energy gained by the stars for each spin configuration (solid curves). Stars are added from the largest specific energy gain downwards, and the horizontal axis is reversed so that this corresponds to moving from left to right, with the high-energy tail on the left. The total reaches its maximum at $\Delta E_*/m_*=0$, once all stars that gained energy have been included, and then decreases slightly as stars that lost energy are added. The horizontal dashed lines give the total energy lost by the binary to ejected stars ($-\Delta E_{\rm bin,eject}$). Comparing the final value of each solid curve with its dashed line therefore shows the net transfer of binding energy from the binary to its stellar environment.

Stars with very large gains ($\Delta E_*/m_*\gtrsim10^{8}\,{\rm km^2\,s^{-2}}$) contribute negligibly. The curves rise steeply towards $\Delta E_*/m_*=0$, showing that most of the energy is deposited by stars with $0\lesssim\Delta E_*/m_*\lesssim5\cdot10^{7}\,{\rm km^2\,s^{-2}}$, and reach a maximum of $(2.9$--$3.1)\cdot10^{14}\,\mathrm{M_\odot}\,{\rm km^2\,s^{-2}}$ at the threshold of $\Delta E_*/m_*=0$, where all stars that gained energy have been included. For $\Delta E_*/m_*<0$ the curves decline, because stars that lost energy to the binary return part of the gain. This reduction amounts to about 15--22\% of the peak value and saturates by $\Delta E_*/m_*\approx-2\cdot10^{7}\,{\rm km^2\,s^{-2}}$, beyond which the curves become flat. The plateau is the net energy transferred from the binary to the stars. For all runs it coincides with the corresponding dashed line to better than $\sim1\%$, which demonstrates energy conservation in the binary--star system and confirms that the encounters in the high-energy tail are responsible for the hardening. 

\begin{figure}[h!]
    \centering   
    \includegraphics[width=0.75\linewidth]{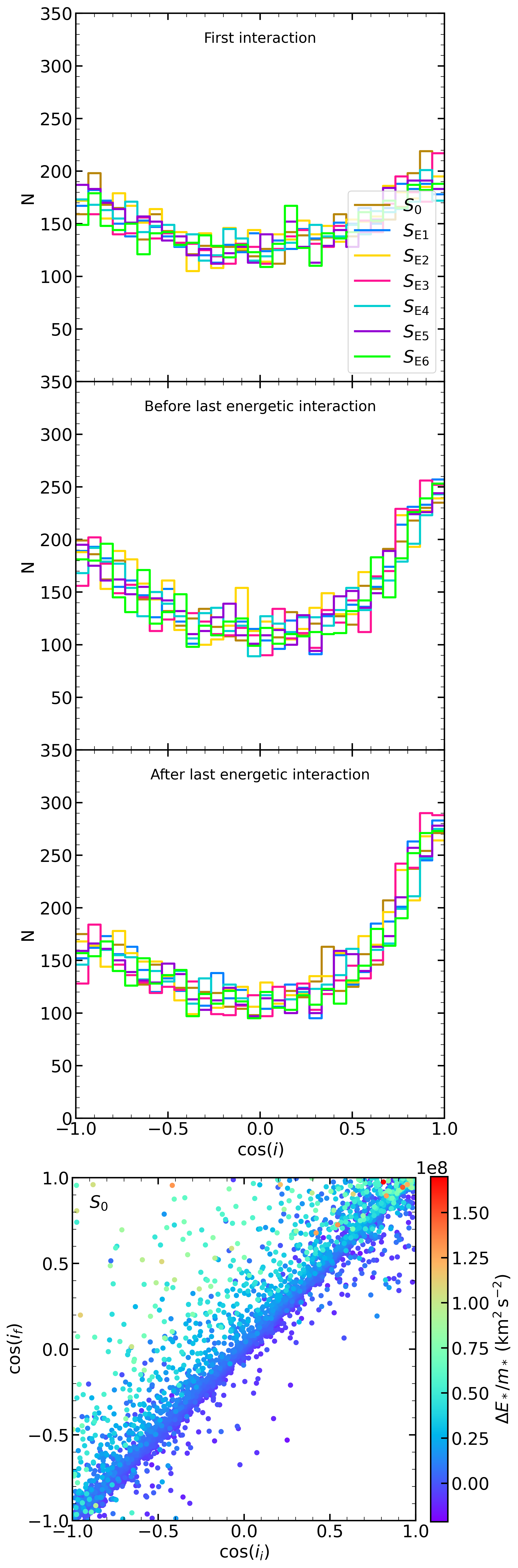}
    \caption{\emph{Top three panels}: Distributions of orbital inclinations of the encounters relative to the black hole binary at the first interaction passage, before the last interaction and after the last interaction for all models respectively. 
    \emph{Bottom} : Two-dimensional histogram between the cosine of the initial ($\cos i_{\rm i}$) and final ($\cos i_{\rm f}$) inclination angles of stellar particles interacting with the binary system color coded by their specific energy change $(\Delta E_*/m_*)$ for model $S_0$.}
    \label{star-balance}
\end{figure}

The differences between spin configurations are small. The plateau values range from $\simeq2.34\cdot10^{14}$ ($S_{\rm E3}$, $S_{\rm E6}$) through $2.39\cdot10^{14}$ ($S_{\rm E4}$) and $\simeq2.48$--$2.50\cdot10^{14}$ ($S_0$, $S_{\rm E2}$) to $\simeq2.52$--$2.53\cdot10^{14}\,\mathrm{M_\odot}\,{\rm km^2\,s^{-2}}$ ($S_{\rm E5}$, $S_{\rm E1}$), i.e.\ a spread of $\lesssim8\%$. This variation does not depend monotonically on spin magnitude or orientation. The half-spin run $S_{\rm E6}$ lies at the lower end of the energy range, as does the full-spin run $S_{\rm E3}$. In contrast, $S_{\rm E1}$, which has twice the spin magnitude of $S_{\rm E6}$, has the maximum energy transfer. Runs $S_{\rm E2}$, $S_{\rm E3}$ and $S_{\rm E4}$ all have in-plane spins, yet $S_{\rm E2}$ groups with the higher-energy runs. The non-spinning model $S_0$ lies within $\sim2\%$ of $S_{\rm E1}$ and inside the range spanned by the spinning models. At the extremes the ordering mirrors Table~\ref{table2} ($S_{\rm E1}$ has the largest plateau, the largest hardening rate and the earliest merger, $S_{\rm E6}$ the smallest plateau, the smallest hardening rate and the latest merger), but the intermediate runs do not follow a monotonic sequence. Since each configuration is a single realization, we cannot separate a genuine spin dependence from the stochasticity of the stellar encounters at this level. We conclude that spin leaves at most a small imprint on how efficiently energy is redistributed to the stars, and hence does not show significantly large differences in the merging timescales.

Once a star comes into close contact with the binary, it exchanges energy and angular momentum with it via three-body interactions. The star may then be kicked out with very high velocity in a single, highly energetic slingshot, or be thrown onto a closed, precessing orbit near the binary and return repeatedly for further close approaches without escape, gaining more energy each time. Such stars may eventually be ejected or captured, but can remain quasi-bound for many orbits. This is the primary driver of the hardening phase, shrinking the binary's separation once it is bound.

During these close interactions, the stars can be classified by the orientation of their orbit into prograde and retrograde populations. Prograde stars orbit in the same direction as the binary's orbital angular momentum ($\cos i\geq0$), so their angular momentum vectors are roughly aligned with that of the binary, while retrograde stars ($\cos i<0$) orbit in the opposite direction and are anti-aligned. We measure the inclination of the interacting stars relative to the binary orbital plane from the orientation of their angular momentum vectors when it first enter, immediately before their final energetic interaction, and immediately after it (top three panels of Fig.~\ref{star-balance}). In all configurations the stars concentrate at $\cos i\to\pm1$, i.e.\ close to the binary orbital plane, with roughly $\approx120$ per bin at $\cos i\approx0$ against $\sim170$--$220$ in the outermost bins at first interaction. The distribution is nearly symmetric at the first interaction, but becomes increasingly skewed towards prograde orbits. The outermost prograde bin reaches roughly $\approx250$ before and 270--290 after the last energetic interaction, compared with roughly $\approx160$--$200$ for the outermost retrograde bin. Prograde stars therefore form the majority of the stars that interact most strongly with the binary, although the excess is modest.

The inclination distribution is not isotropic. The interacting stars show an excess near $\cos i = \pm1$ and a modest prograde excess (Fig.~\ref{star-balance}). Loss-cone selection fixes the magnitude of the angular momentum, not its direction, so it does not by itself favor the binary plane. A prograde asymmetry in the orbital inclinations is expected if the host rotates, since a rotating system contains more stars moving in one sense than the other \citep{2017ApJ...837..135R}. The remnant is flattened, with rotation about an axis inclined to the binary orbit \citep{2021A&A...649A..41A}, but whether the excess in the interacting sample reflects this or the encounters themselves requires the all-stars baseline. In the late phase the population of low angular-momentum stars is sparse, which increases the scatter of the first-interaction histogram. Because all spin runs share the same binary orientation and stellar sample, the inclination distribution is identical across all the models.

The bottom panel of Fig.~\ref{star-balance} (model $S_0$, the other models are shown in Fig.~\ref{fig:All_models_cosi_cosf}) compares the initial and final inclination of each star, color-coded by its energy change. Most stars lie along the diagonal, i.e.\ their orbital orientation is hardly changed by the encounter. The most energetic interactions (orange and red points, $\Delta E_*/m_*\gtrsim10^{8}\,{\rm km^2\,s^{-2}}$) start on prograde orbits and remain nearly aligned with the binary's angular momentum ($\cos i_{\rm f}\to1$). A second population lies above the diagonal at $\cos i_{\rm i}<0$, these are stars that arrive on retrograde orbits and leave on prograde ones, with intermediate energy gains of $\sim(0.5$--$1)\cdot10^{8}\,{\rm km^2\,s^{-2}}$. Both features are consistent with the energy change of a star being proportional to the change of its angular momentum component along the binary axis \citep[Eq.~(63) in][]{2017ApJ...837..135R}.

\begin{figure}[]
    \centering 
    \includegraphics[width=\columnwidth]{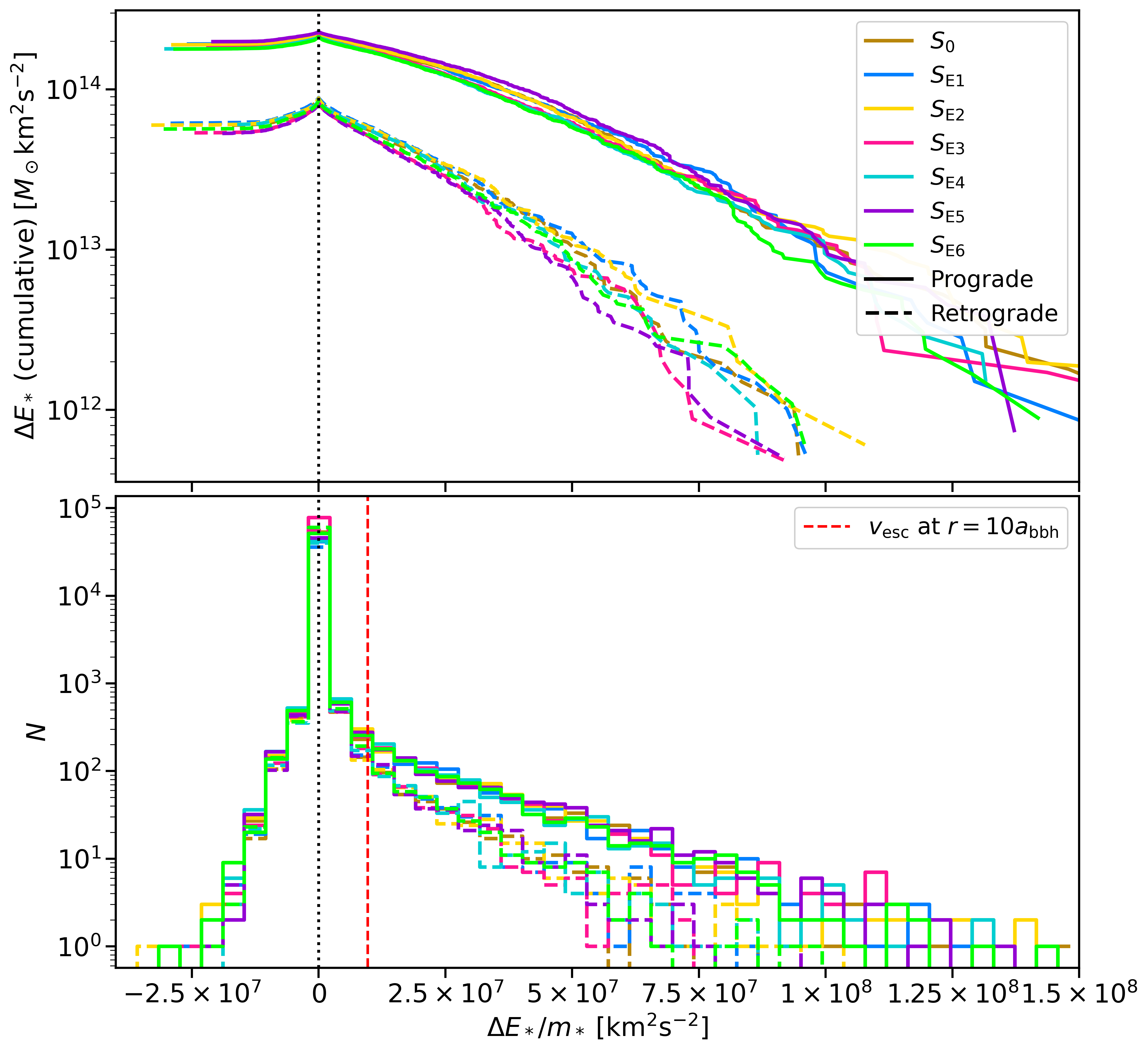}
    \caption{\emph{Top} : Cumulative total energy contributed by stars on prograde and retrograde orbits, shown as a function of their specific energy change, and \emph{Bottom} : Statistics of stellar encounters as a function of specific energy change for the different spin configurations. The dashed red line represents the escape-energy threshold corresponding to the escape speed from the overall galactic system at a radius of 10$a$ from the binary center of mass, at the start of the runs.}
    \label{N-vf}
\end{figure}

Fig.~\ref{N-vf} (top) shows the cumulative energy change of initially prograde ($\cos i_{\rm i}\geq0$, solid) and initially retrograde ($\cos i_{\rm i}<0$, dashed) interactions as a function of the specific energy change, for all configurations. The two curves sum to the total of Fig.~\ref{fig:energybalance}. At $\Delta E_*/m_*=0$ the prograde curves reach $\simeq(2.0$--$2.3)\cdot10^{14}$ and the retrograde ones $\simeq(0.8$--$0.9)\cdot10^{14}\,\mathrm{M_\odot}\,{\rm km^2\,s^{-2}}$, so that prograde encounters carry about three quarters of the energy gained by the stars. The ratio of the two curves grows with energy, from $\approx3$ at $\Delta E_*/m_*=0$ to $\approx4$ at $2.5\cdot10^{7}$, $\approx6$ at $5\cdot10^{7}$ and $\approx10$ at $7.5\cdot10^{7}\,{\rm km^2\,s^{-2}}$. The prograde interactions therefore dominate the high-energy tail in all configurations, spinning as well as the non-spinning. The lower panel shows that the overwhelming majority of the interactions ($\sim10^{5}$ per bin) produce a negligible change in energy, and only a small fraction populate the tail towards large $\Delta E_*/m_*$ (a few hundred per bin above $10^{7}$, and later falling to $\sim25$ at $\sim8\cdot10^{7}$\,${\rm km^2\,s^{-2}}$). At every $\Delta E_*/m_*>0$ the prograde counts exceed the retrograde ones. The dashed red line marks $\Delta E_*/m_*\simeq10^{7}\,{\rm km^2\,s^{-2}}$, corresponding to an equivalent escape speed $\varv_\text{esc}$ $\approx4.4\cdot10^{3}\,{\rm km\,s^{-1}}$ at 10$a$. 
Hence, all the stars to the right of  $\varv_\text{esc}$ are the most high-energy stars which have gained enough specific energy to be candidates for ejection from the system. For more details on the energetic encounters see Appendix \ref{app:duration}.

Prograde stars move in the same direction as the binary’s orbital motion, which leads to longer interaction times, since their velocity relative to the black holes is smaller. For a fixed impact parameter, a smaller relative velocity gives a more nearly parabolic encounter, so the deflection is stronger and the energy exchange larger \citep[see,][]{2021A&A...649A..41A}. They therefore undergo more energetic slingshot encounters, which produce stronger gravitational kicks and higher ejection velocities. 

Prograde stars may also exchange energy and angular momentum more efficiently when their angular-momentum vector is partly aligned with that of the binary. A plausible reason is that the lighter black hole, which lies farther from the COM and moves faster (by a factor $m_1/m_2 \approx 3.75$), is more likely to encounter passing stars and to transfer more energy per encounter, since $\Delta E \sim v\,\Delta p$. We do not separate the contributions of the two black holes here, so this remains a hypothesis. Prograde stars also undergo multiple close encounters or be temporarily trapped in resonant orbits, which cumulatively increases their final kinetic energy before ejection. Retrograde stars, in contrast, typically have higher approach speeds, which leads to single-pass fly-bys and hence less energy exchange.

%%%%%%%%%%%%%%%%%%%%%%%%%%%%%%%%%%%%%%%%%%%%%%%%%%%%%%%%%%%%%%%%%%%

%%%%%%%%%%%%%%%%%%%%%%%%%%%%%%%%%%%%%%%%%%%%%%%%%%%%%%%%%%%%%%%%%%%

\subsection{Size of the stellar system}\label{Size of the stellar system}

An important aspect, already introduced in Sect.~\ref{Initial conditions}, is the number of particles with which the $N$-body runs are performed. The merger simulation of \citet{2016ApJ...828...73K} provides an initial model with $\sim6\cdot10^{6}$ particles (model $S_{\rm H1}$ in Table~\ref{table1}). Because the cost of direct summation scales as $N^{2}$, running all seven spin configurations with this model is very costly, and we therefore reduce it to $1.5\cdot10^{6}$ particles within $r_{\rm cut}=0.1$~kpc (model $S_{\rm E1}$).

To quantify the effect of this reduction on the energy exchange between the binary and the stars, we performed a resolution study with the spin orientation of $S_{\rm E1}$ (models $S_{\rm A1}$--$S_{\rm H1}$ in Table~\ref{table1}), varying $N$ from $2.19\cdot10^{5}$ to $6\cdot10^{6}$. The models are generated by cutting the particle list at a radius $r_{\rm cut}$, from 0.02 to 10~kpc, without altering the remaining particles. As Fig.~\ref{Cum-mass-num-dist} shows, all models therefore have identical mass and number profiles out to their respective $r_{\rm cut}$. Hence, $N$ and the radial extent of the system change together, while the mass resolution stays constant. Inside $\approx100$~pc every particle has a mass of $\approx6\cdot10^{3}\,\mathrm{M_\odot}$, increasing to $\approx10^{4}\,\mathrm{M_\odot}$ at kpc scales. The influence radius of the binary (see Eq.~(\ref{eq:rinf})) is $r_{\rm inf}\approx15$~pc, so the smallest model ( $S_{\rm A1},\ r_{\rm cut}=20$~pc) extends only to $\approx1.4\,r_{\rm inf}$, models $S_{\rm B1}$--$S_{\rm D1}$ to $\approx2$--$3.5\,r_{\rm inf}$, and $S_{\rm E1}$ to $\approx7\,r_{\rm inf}$. 

The particle number is a central concern in direct $N$-body simulations of SMBH binaries because it controls the contribution of two-body relaxation to the supply of stars to the binary. In spherical systems the loss cone is refilled by relaxation alone, and the hardening rate decreases roughly as $N^{-1}$ \citep{2003ApJ...596..860M,2005ApJ...633..680B,2006ApJ...642L..21B}, so that $N$-body results cannot be extrapolated straightforwardly to real galaxies. In non-spherical systems stars can also be supplied by large-scale torques, which do not rely on relaxation, \citet{2006ApJ...642L..21B} found a hardening rate independent of $N$ up to $N=4\cdot10^{5}$ in strongly rotating, bar-unstable models, and \citet{2015ApJ...810...49V} showed with a Monte Carlo method that in the collisionless limit triaxiality alone is sufficient to keep the binary shrinking. However, \citet{2014ApJ...785..163V} found, in $N$-body models of isolated spherical, axisymmetric and triaxial galaxies, a significant $N$ dependence of the hardening rate for $10^{5}\leq N\leq10^{6}$ in all geometries, and concluded that collisional and collisionless effects contribute comparably even at the highest $N$. A fully collisionless regime is therefore not guaranteed at $N\sim10^{6}$, and no sharp threshold in $N$ exists.

\begin{figure}[h]
    \centering 
    \includegraphics[width=\linewidth]{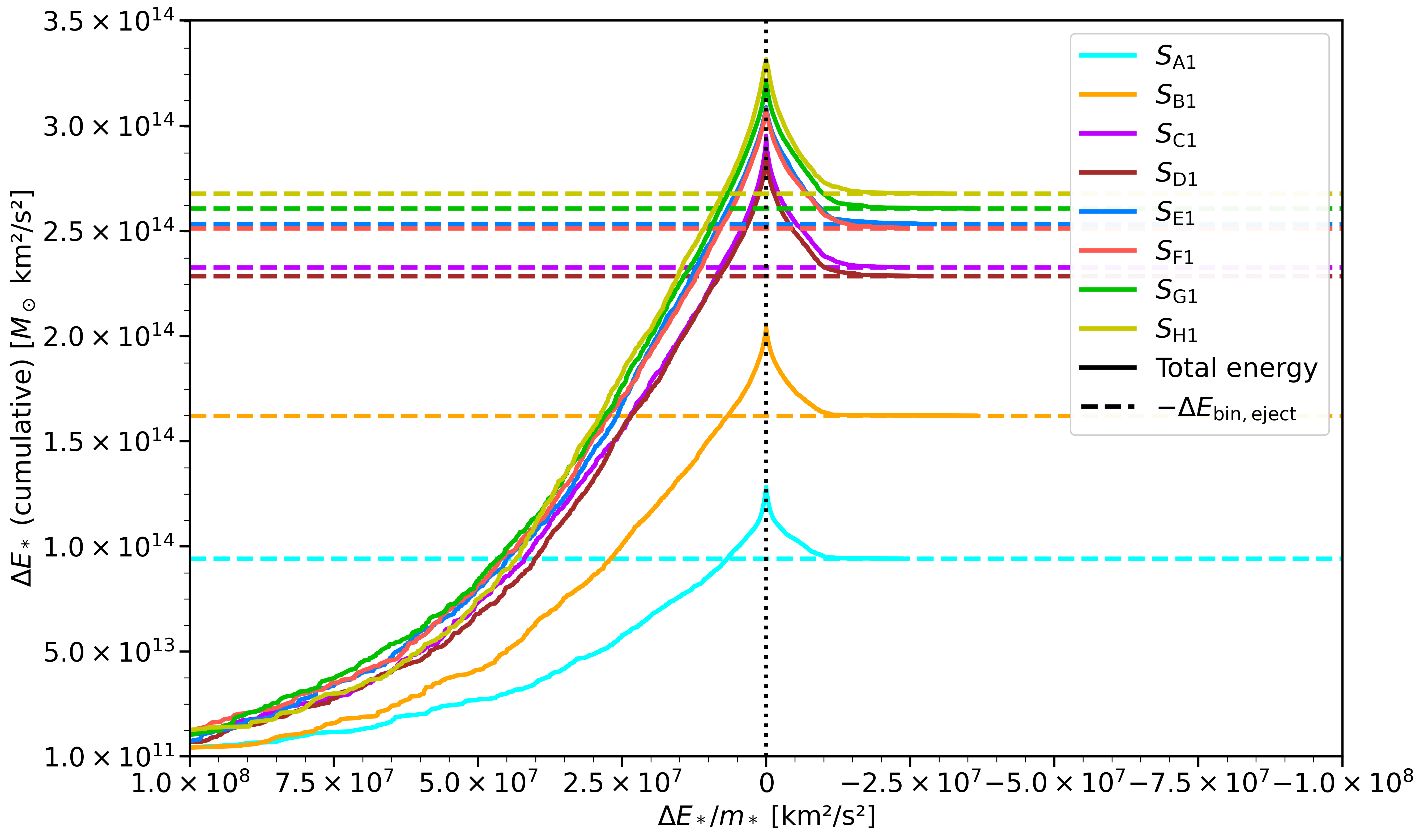}
    \caption{The cumulative energy change of the stellar particles as a function of their specific energy change showing the total energy balance for the spin configuration with different particle numbers. Please note that the x-axis is reversed.}
    \label{diff-N-energy-balance}
\end{figure}

Fig.~\ref{diff-N-energy-balance} shows the energy balance, constructed as in Fig.~\ref{fig:energybalance}, for models $S_{\rm A1}$--$S_{\rm H1}$. The curves have the same shape, but the net energy transferred to the stars (the plateau, equal to $\Delta E_{\rm bin,eject}$) grows with $N$, from $0.94\cdot10^{14}$ for $S_{\rm A1}$ ($N=2.19\cdot10^{5}$) to $2.68\cdot10^{14}\,\mathrm{M_\odot}\,{\rm km^2\,s^{-2}}$ for $S_{\rm H1}$ ($N=6\cdot10^{6}$). The deficit is large for $N\lesssim4\cdot10^{5}$ ($r_{\rm cut}\lesssim30$~pc), where the models recover only $\approx35$--$60\%$ of the energy exchange of $S_{\rm H1}$. Models with $N=(6$--$8)\cdot10^{5}$ lie within $\approx15\%$ of $S_{\rm H1}$, and $S_{\rm E1}$ ($N=1.5\cdot10^{6}$, $2.53\cdot10^{14}$) reaches almost $94\%$. Beyond this the increase is slow such that a factor of four in $N$ between $S_{\rm E1}$ and $S_{\rm H1}$ raises the energy exchange by only $\approx6\%$. The remaining small non-monotonic differences between neighbouring models ($\lesssim2\%$) are smaller than the spread between the spin configurations in Fig.~\ref{fig:energybalance}, and we do not consider them significant.

Since $N$ is varied here by changing the outer radius at fixed mass resolution, Fig.~\ref{diff-N-energy-balance} measures primarily how much of the stellar reservoir around the binary has to be included. Stars initially located between $r_{\rm cut}$ and $\sim100$~pc have orbital times of order $0.1$--$0.2$~Myr, much shorter than the $\approx3.8$~Myr of the runs, so they can be scattered by the binary during the simulation. Excluding them lowers the energy exchange, most strongly for $r_{\rm cut}\lesssim2\,r_{\rm inf}$. Extending the system beyond $100$~pc, in contrast, changes the result by only few per cent. The local relaxation time depends on the particle mass and the local density, which are the same in all models, so this test does not separate the effect of discreteness noise from that of the system size, and a residual collisional contribution to the hardening \citep{2014ApJ...785..163V} cannot be excluded. All spin runs share the same particle realisation and $N$, so such resolution effects act on them in the same way and the comparison between spin configurations is differential. Nevertheless, the $\approx6\%$ difference between $S_{\rm E1}$ and $S_{\rm H1}$ indicates the level to which the absolute energies of Sect.~\ref{Energy Balance of the binaries} are converged, comparable to the spin-dependent spread found there ($\lesssim8\%$). 

We therefore adopt \(N=1.5\cdot10^{6}\) for all spin runs. This choice includes all particles within \(100\)~pc, where the mass resolution is uniform, and reduces the particle number from \(6\cdot10^{6}\) to \(1.5\cdot10^{6}\), lowering the computational cost of direct force evaluation by approximately a factor of 16.
%%%%%%%%%%%%%%%%%%%%%%%%%%%%%%%%%%%%%%%%%%%%%%%%%%%%%%%%%%%%%%

%%%%%%%%%%%%%%%%%%%%%%%%%%%%%%%%%%%%%%%%%%%%%%%%%%%%%%%%%%%%%%

\section{Discussion and conclusions}\label{sec:Conclusions and Discussions}

In this paper, we have investigated the dynamical evolution and merger of a spinning SMBH binary in a dense stellar environment using direct $N$-body simulations with PN corrections up to 2.5PN order between the SMBHs. The initial conditions were taken from \cite{2016ApJ...828...73K} and \cite{2021A&A...649A..41A} derived from the cosmological galaxy merger simulations of \cite{2015MNRAS.446.1939F} and \cite{2015MNRAS.446.1957F} . We systematically explored the impact of different spin magnitudes and mutual orientations with a non-spinning reference, alongside  a resolution study varying the interacting stellar particle number from $N \approx 2.19 \cdot 10^5$ to $6 \cdot 10^6$.

In all runs the binary coalesces within $\approx 3.8$\,Myr after the start of the simulation. For the non-spinning model, the coalescence time lies close to the revised estimate of \citet{2021A&A...649A..41A} for the same merger remnant ($38.3 \pm 0.8$\,Myr since the particle-splitting time), who showed that the shorter estimate of \citet{2016ApJ...828...73K} was affected by insufficient time resolution of the binary. The dense, slightly triaxial remnant therefore still leads to efficient hardening and coalescence well within a Hubble time, even though the timescale is longer than first reported. Our runs begin at $a \approx 0.018$\,pc, after the binary has formed, so they cover only the final phase of that evolution. Earlier estimates of coalescence times often extrapolated a nearly constant hardening rate into the GW-dominated regime \citep{2015ApJ...798..103K, 2015ApJ...810..139H, 2015MNRAS.454L..66S, 2017ApJ...840...53R}. Here the evolution is followed directly to merger with PN dynamics, as in \citet{2018A&A...615A..71K}, who included PN terms up to 3.5PN and a stellar mass function. 

We find no significant dependence of the coalescence time on different spin configurations. All runs merge within a narrow spread of $0.065$\,Myr which is approximately $0.17\%$ of the absolute simulation time, or $\approx 1.7\%$ of the $\approx 3.8$\,Myr inspiral. The non-spinning run lies inside this range and is not the slowest. The hardening rates span $\approx 5\%$, following the merger ordering at the extremes but not in between. Thus, no monotonic trend with spin magnitude or orientation is present. The small spin signature is expected from the timescale hierarchy. The precession time is much longer than the orbital period throughout the run, so the binary completes many orbits per precession cycle \citep{PhysRevD.103.124026}. The spin--orbit and spin--spin couplings \citep{1995PhRvD..52..821K} therefore act as a weak perturbation that accumulates slowly, meaning the precession angle between the orbital angular momentum and the total angular momentum remains very small throughout our runs, further suppressing differential spin signatures. Meanwhile stellar hardening and, subsequently, GW emission \citep{1996NewA....1...35Q} drives most of the evolution. Because Eq.~(\ref{T_prec}) uses the simulated $a$, this timescale ratio is a consistency check and not an independent test of spin effects. Any spin signature is expected to arise mainly in the last phase. Numerical-relativity studies of vacuum binaries show that spins can strongly influence the last orbits before merger \citep{2006PhRvD..74d1501C}. That regime lies at separations of a few $R_{\rm s}$, where a 2.5PN treatment is least accurate, and it contains no stellar environment. In our runs, stellar encounters drive most of the evolution, which explains why the spin signature is small.

The energy balance confirms that the high-energy encounters are responsible for the hardening, as also found by \citet{2021A&A...649A..41A} at an earlier stage of the evolution. Both studies find that prograde stars are more numerous among the strongly interacting stars, and that some retrograde stars change the sign of their angular momentum and leave on prograde orbits \citep[see also][]{2014ApJ...780..164W}. This is consistent with the energy change scaling with the change in the angular-momentum component along the binary axis \citep{2017ApJ...837..135R}. The two studies differ, however, for the highest-energy encounters. \citet{2021A&A...649A..41A} found that these are almost all initially retrograde and that retrograde stars contribute a large fraction of the total energy exchange, whereas in our runs prograde stars dominate the high-energy tail and carry about three quarters of the energy gain, including the non-spinning run, so this is a property of the binary--star interaction and not of spin.  This difference may reflect the later phase of the binary evolution covered in this work, the different treatment of the stellar system, or the way the inclinations are defined, and we leave a detailed comparison for future work. \citet{2021A&A...649A..41A} attribute the larger number of prograde stars to their higher capture probability \citep{2001ApJ...563...34M, 2014ApJ...780..164W}, while the retrograde encounters reach much smaller pericentre distances, which makes them more energetic. We have not tested these explanations in our runs as it lies far beyond the late phase studied in this work.

Similar prograde--retrograde asymmetries have been reported for the binary itself. Counter-rotating binaries in rotating hosts have been found to reverse their sense of rotation around the time of binary formation, which has been attributed to the preferential ejection of prograde stars \citep{2012MNRAS.420L..38G, 2017MNRAS.470..940M}. 

Our adopted number of particles captures most of the energy exchange of the largest run, in line with the finding that about $10^{6}$ particles are needed to follow the binary beyond the hard-binary stage \citep{2015ApJ...810..139H}. In spherical systems, loss-cone refilling by relaxation makes the hardening rate fall roughly as $N^{-1}$ \citep{2003ApJ...596..860M, 2005ApJ...633..680B}. In non-spherical systems, large-scale torques also supply stars \citep{2006ApJ...642L..21B, 2015ApJ...810...49V}, but a significant $N$ dependence persists up to $N \sim 10^{6}$ \citep{2014ApJ...785..163V}. Our test does not separate relaxation from system size, so a residual collisional contribution cannot be excluded. The remaining difference to the largest run is comparable to the spin-dependent spread, which limits how small an effect we could detect.

The eccentricity decreases in all runs. This differs from the rise of the eccentricity in the original run of \citet{2016ApJ...828...73K}, which \citet{2021A&A...649A..41A} attribute to numerical artifacts. In non-rotating nuclei, stellar encounters slowly increase the eccentricity of a hard binary, whereas in rotating nuclei co-rotating binaries are circularized \citep{2010ApJ...719..851S, 2017ApJ...837..135R}. In three-body scattering experiments, retrograde stars extract angular momentum from the binary more efficiently and drive eccentricity growth, whereas prograde stars promote circularisation \citep{2011MNRAS.415L..35S}, this picture was formalised by \citet{2017ApJ...837..135R} and tested with $N$-body models of rotating galaxies \citep{2015ApJ...810..139H, 2017MNRAS.470..940M, 2020MNRAS.492..256K}. The decrease of the eccentricity in our runs, where prograde stars dominate the energy exchange, is qualitatively in line with this. The remnant rotates \citep{2021A&A...649A..41A}, so a decrease of $e_r$ is plausible for a binary that is roughly aligned with this rotation, but we have not verified the alignment in our runs, nor separated the contributions to the eccentricity evolution. We note that the size of the decrease is also consistent with the circularisation expected from GW emission \citep{PhysRev.136.B1224}.

Each configuration is a single realisation, so a genuine spin dependence cannot be separated from the stochasticity of the stellar encounters. At $a/R_{\rm s}$ from $\approx 500$ down to a few tens, the 1.5PN and 2PN spin terms are small compared with stellar hardening. Any spin signature is therefore expected to be small and to arise mainly in the last phase. We did not analyze waveforms, recoil or remnant spin. Our results show no evidence that spin significantly shortens the merger time for this initial condition. Establishing a small spin effect requires ensembles of realisations or controlled experiments with fixed encounter geometry \citep{1996NewA....1...35Q, 2006ApJ...651..392S, 2008AJ....135.2398M}.  

The small precession angles observed throughout our runs indicate that relativistic spin couplings remain a weak perturbation in this stellar environment, further explaining the minimal variation across configurations. To investigate whether star–BH PN interactions play a more prominent role, we are currently working on a companion paper (Paper II) using a regularized integrator for three-body scattering experiments with higher-order PN terms (up to 3.5PN) and comprehensive spin treatments for both black holes and stars. This will enable us to evaluate the spin dependence of binary–star encounters in a more robust, controlled setting

\begin{acknowledgements}
The authors gratefully acknowledge the Gauss Center for Supercomputing eV\footnote{\url{www.gauss-centre.eu}} for funding this project by providing computing time through the John von Neumann Institute for Computing (NIC) on the GCS Supercomputers JUPITER and JUWELS at Jülich Supercomputing Center (JSC). The authors acknowledge support by the High Performance and Cloud Computing Group at the Zentrum für Datenverarbeitung of the University of Tübingen, the state of Baden-Württemberg through bwHPC and the German Research Foundation (DFG) through grant no. INST 37/1159-1 FUGG, and the data storage service SDS@hd supported by the Ministry of Science, Research and the Arts Baden-Württemberg (MWK) and the German Research Foundation (DFG) through grant no. INST 35/1503-1 FUGG. NS acknowledges the support of doctoral scholarship from the German Academic Exchange Service (DAAD) through grant no. 91864951. PB and MS thank the Polish high-performance computing infrastructure PLGrid (HPC Centre: ACK Cyfronet AGH -- docs.hpc.cyfronet.pl) for providing Helios computer facilities and support within computational grant No.~PLG/2026/019243. 

\end{acknowledgements}


%\bibliographystyle{aa} %% only use for submission

\bibliographystyle{aa}  %% use for production and arXiv: produces ULR links in the pdf

\bibliography{references}  

@ARTICLE{Sobolenko2022,
       author = {{Sobolenko}, M. and {Kompaniiets}, O. and {Berczik}, P. and {Marchenko}, V. and {Vasylenko}, A. and {Fedorova}, E. and {Shukirgaliyev}, B.},
        title = "{NGC 6240 supermassive black hole binary dynamical evolution based on Chandra data}",
      journal = {\mnras},
         year = 2022,
        month = dec,
       volume = {517},
       number = {2},
        pages = {1791-1802},
          doi = {10.1093/mnras/stac2472},
archivePrefix = {arXiv},
       eprint = {2209.01264},
 primaryClass = {astro-ph.GA},
       adsurl = {https://ui.adsabs.harvard.edu/abs/2022MNRAS.517.1791S}
}

@ARTICLE{Sobolenko2017,
       author = {{Sobolenko}, M. and {Berczik}, P. and {Spurzem}, R. and {Kupi}, G.},
        title = "{Fast coalescence of post-Newtonian Supermassive Black Hole Binaries in real galaxies}",
      journal = {Kinematics and Physics of Celestial Bodies},
         year = 2017,
        month = jan,
       volume = {33},
       number = {1},
        pages = {21-37},
       adsurl = {https://ui.adsabs.harvard.edu/abs/2017KPCB...33...21S}
}

@ARTICLE{Rantala2018,
       author = {{Rantala}, Antti and {Johansson}, Peter H. and {Naab}, Thorsten and {Thomas}, Jens and {Frigo}, Matteo},
        title = "{The Formation of Extremely Diffuse Galaxy Cores by Merging Supermassive Black Holes}",
      journal = {\apj},
         year = 2018,
        month = sep,
       volume = {864},
       number = {2},
          eid = {113},
        pages = {113},
          doi = {10.3847/1538-4357/aada47},
archivePrefix = {arXiv},
       eprint = {1805.10295},
 primaryClass = {astro-ph.GA},
       adsurl = {https://ui.adsabs.harvard.edu/abs/2018ApJ...864..113R}
}

@ARTICLE{Rantala2019,
       author = {{Rantala}, Antti and {Johansson}, Peter H. and {Naab}, Thorsten and {Thomas}, Jens and {Frigo}, Matteo},
        title = "{The Simultaneous Formation of Cored, Tangentially Biased, and Kinematically Decoupled Centers in Massive Early-type Galaxies}",
      journal = {\apjl},
         year = 2019,
        month = feb,
       volume = {872},
       number = {2},
          eid = {L17},
        pages = {L17},
          doi = {10.3847/2041-8213/ab04b1},
archivePrefix = {arXiv},
       eprint = {1812.02732},
 primaryClass = {astro-ph.GA},
       adsurl = {https://ui.adsabs.harvard.edu/abs/2019ApJ...872L..17R}
}

@ARTICLE{Mannerkoski2019,
       author = {{Mannerkoski}, Matias and {Johansson}, Peter H. and {Pihajoki}, Pauli and {Rantala}, Antti and {Naab}, Thorsten},
        title = "{Gravitational Waves from the Inspiral of Supermassive Black Holes in Galactic-scale Simulations}",
      journal = {\apj},
         year = 2019,
        month = dec,
       volume = {887},
       number = {1},
          eid = {35},
        pages = {35},
          doi = {10.3847/1538-4357/ab52f9},
archivePrefix = {arXiv},
       eprint = {1909.01373},
 primaryClass = {astro-ph.GA},
       adsurl = {https://ui.adsabs.harvard.edu/abs/2019ApJ...887...35M}
}

@ARTICLE{Sedda2019,
       author = {{Arca Sedda}, Manuel and {Berczik}, Peter and {Capuzzo-Dolcetta}, Roberto and {Fragione}, Giacomo and {Sobolenko}, Margaryta and {Spurzem}, Rainer},
        title = "{Supermassive black holes coalescence mediated by massive perturbers: implications for gravitational waves emission and nuclear cluster formation}",
      journal = {\mnras},
         year = 2019,
        month = mar,
       volume = {484},
       number = {1},
        pages = {520-542},
          doi = {10.1093/mnras/sty3458},
archivePrefix = {arXiv},
       eprint = {1712.05810},
 primaryClass = {astro-ph.GA},
       adsurl = {https://ui.adsabs.harvard.edu/abs/2019MNRAS.484..520A}
}

@ARTICLE{Nitadori2008,
       author = {{Nitadori}, Keigo and {Makino}, Junichiro},
        title = "{Sixth- and eighth-order Hermite integrator for N-body simulations}",
      journal = {\na},
         year = "2008",
        month = "Oct",
       volume = {13},
       number = {7},
        pages = {498-507},
          doi = {10.1016/j.newast.2008.01.010},
archivePrefix = {arXiv},
       eprint = {0708.0738},
 primaryClass = {astro-ph},
       adsurl = {https://ui.adsabs.harvard.edu/abs/2008NewA...13..498N}
}

@INPROCEEDINGS{BSW2013,
   author = {{Berczik}, P. and {Spurzem}, R. and {Wang}, L. and {Zhong}, S. and
	{Huang}, S.},
    title = "{Up to 700k GPU cores, Kepler, and the Exascale future for simulations of star clusters around black holes.}",
booktitle = {Third International Conference ''High Performance Computing'', HPC-UA 2013, p. 52-59},
     year = 2013,
archivePrefix = "arXiv",
   eprint = {1312.1789},
 primaryClass = "astro-ph.IM",
    month = oct,
    pages = {52-59},
   adsurl = {http://adsabs.harvard.edu/abs/2013hpc..conf...52B}
}

@misc{Berczik2024ascl,
       author = {{Berczik}, Peter and {Nitadori}, Keigo and {Zhong}, Shiyan and {Spurzem}, Rainer and {Hamada}, Tsuyoshi and {Wang}, Xiaowei and {Berentzen}, Ingo and {Veles}, Alexander and {Ge}, Wei},
        title = "{phi-GPU: Parallel Hermite Integration on GPU}",
 howpublished = {Astrophysics Source Code Library, record ascl:2406.027},
         year = 2024,
        month = jun,
          eid = {ascl:2406.027},
archivePrefix = {ascl},
       eprint = {2406.027},
       adsurl = {https://ui.adsabs.harvard.edu/abs/2024ascl.soft06027B}
}

@INPROCEEDINGS{Berczik2011,
       author = {{Berczik}, Peter and {Nitadori}, Keigo and {Zhong}, Shiyan and
         {Spurzem}, Rainer and {Hamada}, Tsuyoshi and {Wang}, Xiaowei and
         {Berentzen}, Ingo and {Veles}, Alexander and {Ge}, Wei},
        title = "{High performance massively parallel direct N-body simulations on large GPU clusters.}",
    booktitle = {International conference on High Performance Computing},
         year = "2011",
        month = "Oct",
        pages = {8-18},
       adsurl = {https://ui.adsabs.harvard.edu/abs/2011hpc..conf....8B}
}

@ARTICLE{Khan2026A&A,
       author = {{Khan}, Fazeel Mahmood and {Berczik}, Peter and {Sobolenko}, Margaryta and {Just}, Andreas and {Spurzem}, Rainer and {Holley-Bockelmann}, Kelly and {Macci{\`o}}, Andrea V.},
        title = "{Intermediate-mass black hole binary evolution in nuclear star clusters: The effect of the stellar-mass black hole population}",
      journal = {\aap},
         year = 2026,
        month = feb,
       volume = {706},
          eid = {A354},
        pages = {A354},
          doi = {10.1051/0004-6361/202557283},
archivePrefix = {arXiv},
       eprint = {2512.11038},
 primaryClass = {astro-ph.GA},
       adsurl = {https://ui.adsabs.harvard.edu/abs/2026A&A...706A.354K}
}

@ARTICLE{Berczik2024A&A,
       author = {{Berczik}, P. and {Sobolenko}, M. and {Ishchenko}, M.},
        title = "{Dynamical evolution timescales for the triple supermassive black hole system in NGC 6240}",
      journal = {\aap},
         year = 2024,
        month = jul,
       volume = {687},
          eid = {L18},
        pages = {L18},
          doi = {10.1051/0004-6361/202450141},
archivePrefix = {arXiv},
       eprint = {2407.00473},
 primaryClass = {astro-ph.GA},
       adsurl = {https://ui.adsabs.harvard.edu/abs/2024A&A...687L..18B}
}

@ARTICLE{Koehn2023A&A,
       author = {{Koehn}, H. and {Just}, A. and {Berczik}, P. and {Tremmel}, M.},
        title = "{Dynamics of supermassive black hole triples in the ROMULUS25 cosmological simulation}",
      journal = {\aap},
         year = 2023,
        month = oct,
       volume = {678},
          eid = {A11},
        pages = {A11},
          doi = {10.1051/0004-6361/202347093},
archivePrefix = {arXiv},
       eprint = {2308.10894},
 primaryClass = {astro-ph.GA},
       adsurl = {https://ui.adsabs.harvard.edu/abs/2023A&A...678A..11K}
}

@ARTICLE{Khan2018ApJ,
       author = {{Khan}, Fazeel M. and {Capelo}, Pedro R. and {Mayer}, Lucio and {Berczik}, Peter},
        title = "{Dynamical Evolution and Merger Timescales of LISA Massive Black Hole Binaries in Disk Galaxy Mergers}",
      journal = {\apj},
         year = 2018,
        month = dec,
       volume = {868},
       number = {2},
          eid = {97},
        pages = {97},
          doi = {10.3847/1538-4357/aae77b},
archivePrefix = {arXiv},
       eprint = {1807.11004},
 primaryClass = {astro-ph.GA},
       adsurl = {https://ui.adsabs.harvard.edu/abs/2018ApJ...868...97K}
}

@ARTICLE{2013ARA&A..51..511K,
       author = {{Kormendy}, John and {Ho}, Luis C.},
        title = "{Coevolution (Or Not) of Supermassive Black Holes and Host Galaxies}",
      journal = {\araa},
         year = 2013,
        month = aug,
       volume = {51},
       number = {1},
        pages = {511-653},
          doi = {10.1146/annurev-astro-082708-101811},
archivePrefix = {arXiv},
       eprint = {1304.7762},
 primaryClass = {astro-ph.CO},
       adsurl = {https://ui.adsabs.harvard.edu/abs/2013ARA&A..51..511K}
}

@ARTICLE{2000ApJ...539L...9F,
       author = {{Ferrarese}, Laura and {Merritt}, David},
        title = "{A Fundamental Relation between Supermassive Black Holes and Their Host Galaxies}",
      journal = {\apjl},
         year = 2000,
        month = aug,
       volume = {539},
       number = {1},
        pages = {L9-L12},
          doi = {10.1086/312838},
archivePrefix = {arXiv},
       eprint = {astro-ph/0006053},
 primaryClass = {astro-ph},
       adsurl = {https://ui.adsabs.harvard.edu/abs/2000ApJ...539L...9F}
}

@ARTICLE{1998A&A...331L...1S,
       author = {{Silk}, Joseph and {Rees}, Martin J.},
        title = "{Quasars and galaxy formation}",
      journal = {\aap},
         year = 1998,
        month = mar,
       volume = {331},
        pages = {L1-L4},
          doi = {10.48550/arXiv.astro-ph/9801013},
archivePrefix = {arXiv},
       eprint = {astro-ph/9801013},
 primaryClass = {astro-ph},
       adsurl = {https://ui.adsabs.harvard.edu/abs/1998A&A...331L...1S}
}

@ARTICLE{2006ApJS..163....1H,
       author = {{Hopkins}, Philip F. and {Hernquist}, Lars and {Cox}, Thomas J. and {Di Matteo}, Tiziana and {Robertson}, Brant and {Springel}, Volker},
        title = "{A Unified, Merger-driven Model of the Origin of Starbursts, Quasars, the Cosmic X-Ray Background, Supermassive Black Holes, and Galaxy Spheroids}",
      journal = {\apjs},
         year = 2006,
        month = mar,
       volume = {163},
       number = {1},
        pages = {1-49},
          doi = {10.1086/499298},
archivePrefix = {arXiv},
       eprint = {astro-ph/0506398},
 primaryClass = {astro-ph},
       adsurl = {https://ui.adsabs.harvard.edu/abs/2006ApJS..163....1H}
}

@ARTICLE{2010A&ARv..18..279V,
       author = {{Volonteri}, Marta},
        title = "{Formation of supermassive black holes}",
      journal = {\aapr},
         year = 2010,
        month = jul,
       volume = {18},
       number = {3},
        pages = {279-315},
          doi = {10.1007/s00159-010-0029-x},
archivePrefix = {arXiv},
       eprint = {1003.4404},
 primaryClass = {astro-ph.CO},
       adsurl = {https://ui.adsabs.harvard.edu/abs/2010A&ARv..18..279V}
}

@ARTICLE{2005Natur.433..604D,
       author = {{Di Matteo}, Tiziana and {Springel}, Volker and {Hernquist}, Lars},
        title = "{Energy input from quasars regulates the growth and activity of black holes and their host galaxies}",
      journal = {\nat},
         year = 2005,
        month = feb,
       volume = {433},
       number = {7026},
        pages = {604-607},
          doi = {10.1038/nature03335},
archivePrefix = {arXiv},
       eprint = {astro-ph/0502199},
 primaryClass = {astro-ph},
       adsurl = {https://ui.adsabs.harvard.edu/abs/2005Natur.433..604D}
}

@ARTICLE{2005MNRAS.361..776S,
       author = {{Springel}, Volker and {Di Matteo}, Tiziana and {Hernquist}, Lars},
        title = "{Modelling feedback from stars and black holes in galaxy mergers}",
      journal = {\mnras},
         year = 2005,
        month = aug,
       volume = {361},
       number = {3},
        pages = {776-794},
          doi = {10.1111/j.1365-2966.2005.09238.x},
archivePrefix = {arXiv},
       eprint = {astro-ph/0411108},
 primaryClass = {astro-ph},
       adsurl = {https://ui.adsabs.harvard.edu/abs/2005MNRAS.361..776S}
}

@ARTICLE{1980Natur.287..307B,
       author = {{Begelman}, M.~C. and {Blandford}, R.~D. and {Rees}, M.~J.},
        title = "{Massive black hole binaries in active galactic nuclei}",
      journal = {\nat},
         year = 1980,
        month = sep,
       volume = {287},
       number = {5780},
        pages = {307-309},
          doi = {10.1038/287307a0},
       adsurl = {https://ui.adsabs.harvard.edu/abs/1980Natur.287..307B}
}

@ARTICLE{2007Sci...316.1874M,
       author = {{Mayer}, L. and {Kazantzidis}, S. and {Madau}, P. and {Colpi}, M. and {Quinn}, T. and {Wadsley}, J.},
        title = "{Rapid Formation of Supermassive Black Hole Binaries in Galaxy Mergers with Gas}",
      journal = {Science},
         year = 2007,
        month = jun,
       volume = {316},
       number = {5833},
        pages = {1874},
          doi = {10.1126/science.1141858},
archivePrefix = {arXiv},
       eprint = {0706.1562},
 primaryClass = {astro-ph},
       adsurl = {https://ui.adsabs.harvard.edu/abs/2007Sci...316.1874M}
}

@ARTICLE{1943ApJ....97..255C,
       author = {{Chandrasekhar}, S.},
        title = "{Dynamical Friction. I. General Considerations: the Coefficient of Dynamical Friction.}",
      journal = {\apj},
         year = 1943,
        month = mar,
       volume = {97},
        pages = {255},
          doi = {10.1086/144517},
       adsurl = {https://ui.adsabs.harvard.edu/abs/1943ApJ....97..255C}
}

@ARTICLE{2001ApJ...563...34M,
       author = {{Milosavljevi{\'c}}, Milo{\v{s}} and {Merritt}, David},
        title = "{Formation of Galactic Nuclei}",
      journal = {\apj},
         year = 2001,
        month = dec,
       volume = {563},
       number = {1},
        pages = {34-62},
          doi = {10.1086/323830},
archivePrefix = {arXiv},
       eprint = {astro-ph/0103350},
 primaryClass = {astro-ph},
       adsurl = {https://ui.adsabs.harvard.edu/abs/2001ApJ...563...34M}
}

@ARTICLE{1996NewA....1...35Q,
       author = {{Quinlan}, Gerald D.},
        title = "{The dynamical evolution of massive black hole binaries I. Hardening in a fixed stellar background}",
      journal = {\na},
         year = 1996,
        month = jul,
       volume = {1},
       number = {1},
        pages = {35-56},
          doi = {10.1016/S1384-1076(96)00003-6},
archivePrefix = {arXiv},
       eprint = {astro-ph/9601092},
 primaryClass = {astro-ph},
       adsurl = {https://ui.adsabs.harvard.edu/abs/1996NewA....1...35Q}
}

@ARTICLE{2004ApJ...606..788M,
       author = {{Merritt}, David and {Poon}, M.~Y.},
        title = "{Chaotic Loss Cones and Black Hole Fueling}",
      journal = {\apj},
         year = 2004,
        month = may,
       volume = {606},
       number = {2},
        pages = {788-798},
          doi = {10.1086/382497},
archivePrefix = {arXiv},
       eprint = {astro-ph/0302296},
 primaryClass = {astro-ph},
       adsurl = {https://ui.adsabs.harvard.edu/abs/2004ApJ...606..788M}
}

@ARTICLE{2006ApJ...642L..21B,
       author = {{Berczik}, Peter and {Merritt}, David and {Spurzem}, Rainer and {Bischof}, Hans-Peter},
        title = "{Efficient Merger of Binary Supermassive Black Holes in Nonaxisymmetric Galaxies}",
      journal = {\apjl},
         year = 2006,
        month = may,
       volume = {642},
       number = {1},
        pages = {L21-L24},
          doi = {10.1086/504426},
archivePrefix = {arXiv},
       eprint = {astro-ph/0601698},
 primaryClass = {astro-ph},
       adsurl = {https://ui.adsabs.harvard.edu/abs/2006ApJ...642L..21B}
}

@ARTICLE{2003ApJ...596..860M,
       author = {{Milosavljevi{\'c}}, Milo{\v{s}} and {Merritt}, David},
        title = "{Long-Term Evolution of Massive Black Hole Binaries}",
      journal = {\apj},
         year = 2003,
        month = oct,
       volume = {596},
       number = {2},
        pages = {860-878},
          doi = {10.1086/378086},
archivePrefix = {arXiv},
       eprint = {astro-ph/0212459},
 primaryClass = {astro-ph},
       adsurl = {https://ui.adsabs.harvard.edu/abs/2003ApJ...596..860M}
}

@ARTICLE{2011ApJ...732...89K,
       author = {{Khan}, Fazeel Mahmood and {Just}, Andreas and {Merritt}, David},
        title = "{Efficient Merger of Binary Supermassive Black Holes in Merging Galaxies}",
      journal = {\apj},
         year = 2011,
        month = may,
       volume = {732},
       number = {2},
          eid = {89},
        pages = {89},
          doi = {10.1088/0004-637X/732/2/89},
archivePrefix = {arXiv},
       eprint = {1103.0272},
 primaryClass = {astro-ph.CO},
       adsurl = {https://ui.adsabs.harvard.edu/abs/2011ApJ...732...89K}
}

@ARTICLE{2011ApJ...732L..26P,
       author = {{Preto}, Miguel and {Berentzen}, Ingo and {Berczik}, Peter and {Spurzem}, Rainer},
        title = "{Fast Coalescence of Massive Black Hole Binaries from Mergers of Galactic Nuclei: Implications for Low-frequency Gravitational-wave Astrophysics}",
      journal = {\apjl},
         year = 2011,
        month = may,
       volume = {732},
       number = {2},
          eid = {L26},
        pages = {L26},
          doi = {10.1088/2041-8205/732/2/L26},
archivePrefix = {arXiv},
       eprint = {1102.4855},
 primaryClass = {astro-ph.GA},
       adsurl = {https://ui.adsabs.harvard.edu/abs/2011ApJ...732L..26P}
}

@ARTICLE{2017ApJ...840...53R,
       author = {{Rantala}, Antti and {Pihajoki}, Pauli and {Johansson}, Peter H. and {Naab}, Thorsten and {Lah{\'e}n}, Natalia and {Sawala}, Till},
        title = "{Post-Newtonian Dynamical Modeling of Supermassive Black Holes in Galactic-scale Simulations}",
      journal = {\apj},
         year = 2017,
        month = may,
       volume = {840},
       number = {1},
          eid = {53},
        pages = {53},
          doi = {10.3847/1538-4357/aa6d65},
archivePrefix = {arXiv},
       eprint = {1611.07028},
 primaryClass = {astro-ph.GA},
       adsurl = {https://ui.adsabs.harvard.edu/abs/2017ApJ...840...53R}
}

@article{PhysRev.136.B1224,
  title = {Gravitational Radiation and the Motion of Two Point Masses},
  author = {Peters, P. C.},
  journal = {Phys. Rev.},
  volume = {136},
  issue = {4B},
  pages = {B1224--B1232},
  numpages = {0},
  year = {1964},
  month = {Nov},
  publisher = {American Physical Society},
  doi = {10.1103/PhysRev.136.B1224},
  url = {https://link.aps.org/doi/10.1103/PhysRev.136.B1224}
}

@ARTICLE{2019MNRAS.486.4044B,
       author = {{Bonetti}, Matteo and {Sesana}, Alberto and {Haardt}, Francesco and {Barausse}, Enrico and {Colpi}, Monica},
        title = "{Post-Newtonian evolution of massive black hole triplets in galactic nuclei - IV. Implications for LISA}",
      journal = {\mnras},
         year = 2019,
        month = jul,
       volume = {486},
       number = {3},
        pages = {4044-4060},
          doi = {10.1093/mnras/stz903},
archivePrefix = {arXiv},
       eprint = {1812.01011},
 primaryClass = {astro-ph.GA},
       adsurl = {https://ui.adsabs.harvard.edu/abs/2019MNRAS.486.4044B}
}

@ARTICLE{2004PhRvD..69h2005B,
       author = {{Barack}, Leor and {Cutler}, Curt},
        title = "{LISA capture sources: Approximate waveforms, signal-to-noise ratios, and parameter estimation accuracy}",
      journal = {\prd},
         year = 2004,
        month = apr,
       volume = {69},
       number = {8},
          eid = {082005},
        pages = {082005},
          doi = {10.1103/PhysRevD.69.082005},
archivePrefix = {arXiv},
       eprint = {gr-qc/0310125},
 primaryClass = {gr-qc},
       adsurl = {https://ui.adsabs.harvard.edu/abs/2004PhRvD..69h2005B}
}

@ARTICLE{2004ApJ...611..623S,
       author = {{Sesana}, Alberto and {Haardt}, Francesco and {Madau}, Piero and {Volonteri}, Marta},
        title = "{Low-Frequency Gravitational Radiation from Coalescing Massive Black Hole Binaries in Hierarchical Cosmologies}",
      journal = {\apj},
         year = 2004,
        month = aug,
       volume = {611},
       number = {2},
        pages = {623-632},
          doi = {10.1086/422185},
archivePrefix = {arXiv},
       eprint = {astro-ph/0401543},
 primaryClass = {astro-ph},
       adsurl = {https://ui.adsabs.harvard.edu/abs/2004ApJ...611..623S}
}

@ARTICLE{2017arXiv170200786A,
       author = {{Amaro-Seoane}, Pau and {Audley}, Heather and {Babak}, Stanislav and {Baker}, John and {Barausse}, Enrico and {Bender}, Peter and {Berti}, Emanuele and {Binetruy}, Pierre and {Born}, Michael and {Bortoluzzi}, Daniele and {Camp}, Jordan and {Caprini}, Chiara and {Cardoso}, Vitor and {Colpi}, Monica and {Conklin}, John and {Cornish}, Neil and {Cutler}, Curt and {Danzmann}, Karsten and {Dolesi}, Rita and {Ferraioli}, Luigi and {Ferroni}, Valerio and {Fitzsimons}, Ewan and {Gair}, Jonathan and {Gesa Bote}, Lluis and {Giardini}, Domenico and {Gibert}, Ferran and {Grimani}, Catia and {Halloin}, Hubert and {Heinzel}, Gerhard and {Hertog}, Thomas and {Hewitson}, Martin and {Holley-Bockelmann}, Kelly and {Hollington}, Daniel and {Hueller}, Mauro and {Inchauspe}, Henri and {Jetzer}, Philippe and {Karnesis}, Nikos and {Killow}, Christian and {Klein}, Antoine and {Klipstein}, Bill and {Korsakova}, Natalia and {Larson}, Shane L and {Livas}, Jeffrey and {Lloro}, Ivan and {Man}, Nary and {Mance}, Davor and {Martino}, Joseph and {Mateos}, Ignacio and {McKenzie}, Kirk and {McWilliams}, Sean T and {Miller}, Cole and {Mueller}, Guido and {Nardini}, Germano and {Nelemans}, Gijs and {Nofrarias}, Miquel and {Petiteau}, Antoine and {Pivato}, Paolo and {Plagnol}, Eric and {Porter}, Ed and {Reiche}, Jens and {Robertson}, David and {Robertson}, Norna and {Rossi}, Elena and {Russano}, Giuliana and {Schutz}, Bernard and {Sesana}, Alberto and {Shoemaker}, David and {Slutsky}, Jacob and {Sopuerta}, Carlos F. and {Sumner}, Tim and {Tamanini}, Nicola and {Thorpe}, Ira and {Troebs}, Michael and {Vallisneri}, Michele and {Vecchio}, Alberto and {Vetrugno}, Daniele and {Vitale}, Stefano and {Volonteri}, Marta and {Wanner}, Gudrun and {Ward}, Harry and {Wass}, Peter and {Weber}, William and {Ziemer}, John and {Zweifel}, Peter},
        title = "{Laser Interferometer Space Antenna}",
      journal = {arXiv e-prints},
         year = 2017,
        month = feb,
          eid = {arXiv:1702.00786},
        pages = {arXiv:1702.00786},
          doi = {10.48550/arXiv.1702.00786},
archivePrefix = {arXiv},
       eprint = {1702.00786},
 primaryClass = {astro-ph.IM},
       adsurl = {https://ui.adsabs.harvard.edu/abs/2017arXiv170200786A}
}

@ARTICLE{2014Natur.509..177V,
       author = {{Vogelsberger}, M. and {Genel}, S. and {Springel}, V. and {Torrey}, P. and {Sijacki}, D. and {Xu}, D. and {Snyder}, G. and {Bird}, S. and {Nelson}, D. and {Hernquist}, L.},
        title = "{Properties of galaxies reproduced by a hydrodynamic simulation}",
      journal = {\nat},
         year = 2014,
        month = may,
       volume = {509},
       number = {7499},
        pages = {177-182},
          doi = {10.1038/nature13316},
archivePrefix = {arXiv},
       eprint = {1405.1418},
 primaryClass = {astro-ph.CO},
       adsurl = {https://ui.adsabs.harvard.edu/abs/2014Natur.509..177V}
}

@ARTICLE{2018MNRAS.475..676S,
       author = {{Springel}, Volker and {Pakmor}, R{\"u}diger and {Pillepich}, Annalisa and {Weinberger}, Rainer and {Nelson}, Dylan and {Hernquist}, Lars and {Vogelsberger}, Mark and {Genel}, Shy and {Torrey}, Paul and {Marinacci}, Federico and {Naiman}, Jill},
        title = "{First results from the IllustrisTNG simulations: matter and galaxy clustering}",
      journal = {\mnras},
         year = 2018,
        month = mar,
       volume = {475},
       number = {1},
        pages = {676-698},
          doi = {10.1093/mnras/stx3304},
archivePrefix = {arXiv},
       eprint = {1707.03397},
 primaryClass = {astro-ph.GA},
       adsurl = {https://ui.adsabs.harvard.edu/abs/2018MNRAS.475..676S}
}

@ARTICLE{2017MNRAS.464.3131K,
       author = {{Kelley}, Luke Zoltan and {Blecha}, Laura and {Hernquist}, Lars},
        title = "{Massive black hole binary mergers in dynamical galactic environments}",
      journal = {\mnras},
         year = 2017,
        month = jan,
       volume = {464},
       number = {3},
        pages = {3131-3157},
          doi = {10.1093/mnras/stw2452},
archivePrefix = {arXiv},
       eprint = {1606.01900},
 primaryClass = {astro-ph.HE},
       adsurl = {https://ui.adsabs.harvard.edu/abs/2017MNRAS.464.3131K}
}

@ARTICLE{2019MNRAS.483.3336T,
       author = {{Tremmel}, M. and {Quinn}, T.~R. and {Ricarte}, A. and {Babul}, A. and {Chadayammuri}, U. and {Natarajan}, P. and {Nagai}, D. and {Pontzen}, A. and {Volonteri}, M.},
        title = "{Introducing ROMULUSC: a cosmological simulation of a galaxy cluster with an unprecedented resolution}",
      journal = {\mnras},
         year = 2019,
        month = mar,
       volume = {483},
       number = {3},
        pages = {3336-3362},
          doi = {10.1093/mnras/sty3336},
archivePrefix = {arXiv},
       eprint = {1806.01282},
 primaryClass = {astro-ph.GA},
       adsurl = {https://ui.adsabs.harvard.edu/abs/2019MNRAS.483.3336T}
}

@article{10.1093/mnrasl/slz091,
    author = {Pfister, Hugo and Bar-Or, Ben and Volonteri, Marta and Dubois, Yohan and Capelo, Pedro R},
    title = {Tidal disruption event rates in galaxy merger remnants},
    journal = {Monthly Notices of the Royal Astronomical Society: Letters},
    volume = {488},
    number = {1},
    pages = {L29-L34},
    year = {2019},
    month = {06},
    issn = {1745-3925},
    doi = {10.1093/mnrasl/slz091},
    url = {https://doi.org/10.1093/mnrasl/slz091},
    eprint = {https://academic.oup.com/mnrasl/article-pdf/488/1/L29/56978258/mnrasl\_488\_1\_l29.pdf},
}

@ARTICLE{2017MNRAS.471.4508K,
       author = {{Kelley}, Luke Zoltan and {Blecha}, Laura and {Hernquist}, Lars and {Sesana}, Alberto and {Taylor}, Stephen R.},
        title = "{The gravitational wave background from massive black hole binaries in Illustris: spectral features and time to detection with pulsar timing arrays}",
      journal = {\mnras},
         year = 2017,
        month = nov,
       volume = {471},
       number = {4},
        pages = {4508-4526},
          doi = {10.1093/mnras/stx1638},
archivePrefix = {arXiv},
       eprint = {1702.02180},
 primaryClass = {astro-ph.HE},
       adsurl = {https://ui.adsabs.harvard.edu/abs/2017MNRAS.471.4508K}
}

@ARTICLE{2004PhRvD..70l4020S,
       author = {{Schnittman}, Jeremy D.},
        title = "{Spin-orbit resonance and the evolution of compact binary systems}",
      journal = {\prd},
         year = 2004,
        month = dec,
       volume = {70},
       number = {12},
          eid = {124020},
        pages = {124020},
          doi = {10.1103/PhysRevD.70.124020},
archivePrefix = {arXiv},
       eprint = {astro-ph/0409174},
 primaryClass = {astro-ph},
       adsurl = {https://ui.adsabs.harvard.edu/abs/2004PhRvD..70l4020S}
}

@ARTICLE{2005ApJ...633..680B,
       author = {{Berczik}, Peter and {Merritt}, David and {Spurzem}, Rainer},
        title = "{Long-Term Evolution of Massive Black Hole Binaries. II. Binary Evolution in Low-Density Galaxies}",
      journal = {\apj},
         year = 2005,
        month = nov,
       volume = {633},
       number = {2},
        pages = {680-687},
          doi = {10.1086/491598},
archivePrefix = {arXiv},
       eprint = {astro-ph/0507260},
 primaryClass = {astro-ph},
       adsurl = {https://ui.adsabs.harvard.edu/abs/2005ApJ...633..680B}
}

@ARTICLE{2012ApJ...756...30K,
       author = {{Khan}, Fazeel Mahmood and {Berentzen}, Ingo and {Berczik}, Peter and {Just}, Andreas and {Mayer}, Lucio and {Nitadori}, Keigo and {Callegari}, Simone},
        title = "{Formation and Hardening of Supermassive Black Hole Binaries in Minor Mergers of Disk Galaxies}",
      journal = {\apj},
         year = 2012,
        month = sep,
       volume = {756},
       number = {1},
          eid = {30},
        pages = {30},
          doi = {10.1088/0004-637X/756/1/30},
archivePrefix = {arXiv},
       eprint = {1203.1623},
 primaryClass = {astro-ph.CO},
       adsurl = {https://ui.adsabs.harvard.edu/abs/2012ApJ...756...30K}
}

@INPROCEEDINGS{2012ASPC..453..223S,
       author = {{Spurzem}, R. and {Berczik}, P. and {Zhong}, S. and {Nitadori}, K. and {Hamada}, T. and {Berentzen}, I. and {Veles}, A.},
        title = "{Supermassive Black Hole Binaries in High Performance Massively Parallel Direct N-body Simulations on Large GPU Clusters}",
    booktitle = {Advances in Computational Astrophysics: Methods, Tools, and Outcome},
         year = 2012,
       editor = {{Capuzzo-Dolcetta}, R. and {Limongi}, M. and {Tornamb{\`e}}, A.},
       series = {Astronomical Society of the Pacific Conference Series},
       volume = {453},
        month = jul,
        pages = {223},
       adsurl = {https://ui.adsabs.harvard.edu/abs/2012ASPC..453..223S}
}

@ARTICLE{2021A&A...652A.134S,
       author = {{Sobolenko}, M. and {Berczik}, P. and {Spurzem}, R.},
        title = "{Merging timescale for the supermassive black hole binary in interacting galaxy NGC 6240}",
      journal = {\aap},
         year = 2021,
        month = aug,
       volume = {652},
          eid = {A134},
        pages = {A134},
          doi = {10.1051/0004-6361/202039859},
archivePrefix = {arXiv},
       eprint = {2105.14503},
 primaryClass = {astro-ph.GA},
       adsurl = {https://ui.adsabs.harvard.edu/abs/2021A&A...652A.134S}
}

@INPROCEEDINGS{2016IAUS..312...92V,
       author = {{Vasiliev}, Eugene},
        title = "{Evolution of binary supermassive black holes and the final-parsec problem}",
    booktitle = {Star Clusters and Black Holes in Galaxies across Cosmic Time},
         year = 2016,
       editor = {{Meiron}, Y. and {Li}, S. and {Liu}, F. -K. and {Spurzem}, R.},
       series = {IAU Symposium},
       volume = {312},
        month = feb,
        pages = {92-100},
          doi = {10.1017/S1743921315007607},
archivePrefix = {arXiv},
       eprint = {1411.1762},
 primaryClass = {astro-ph.GA},
       adsurl = {https://ui.adsabs.harvard.edu/abs/2016IAUS..312...92V}
}

@ARTICLE{2022A&A...665A..86B,
       author = {{Berczik}, Peter and {Arca Sedda}, Manuel and {Sobolenko}, Margaryta and {Ishchenko}, Marina and {Sobodar}, Olexander and {Spurzem}, Rainer},
        title = "{Merging of unequal mass binary black holes in non-axisymmetric galactic nuclei}",
      journal = {\aap},
         year = 2022,
        month = sep,
       volume = {665},
          eid = {A86},
        pages = {A86},
          doi = {10.1051/0004-6361/202244354},
archivePrefix = {arXiv},
       eprint = {2008.04342},
 primaryClass = {astro-ph.GA},
       adsurl = {https://ui.adsabs.harvard.edu/abs/2022A&A...665A..86B}
}

@article{PhysRevD.71.129902,
  title = {Erratum: Gravitational-wave inspiral of compact binary systems to $7/2$ post-Newtonian order [Phys. Rev. D 65, 061501(R) (2002)]},
  author = {Blanchet, Luc and Faye, Guillaume and Iyer, Bala R. and Joguet, Benoit},
  journal = {Phys. Rev. D},
  volume = {71},
  issue = {12},
  pages = {129902},
  numpages = {2},
  year = {2005},
  month = {Jun},
  publisher = {American Physical Society},
  doi = {10.1103/PhysRevD.71.129902},
  url = {https://link.aps.org/doi/10.1103/PhysRevD.71.129902}
}

@ARTICLE{2002PhRvD..65h4011G,
       author = {{Gopakumar}, A. and {Iyer}, Bala R.},
        title = "{Second post-Newtonian gravitational wave polarizations for compact binaries in elliptical orbits}",
      journal = {\prd},
         year = 2002,
        month = apr,
       volume = {65},
       number = {8},
          eid = {084011},
        pages = {084011},
          doi = {10.1103/PhysRevD.65.084011},
archivePrefix = {arXiv},
       eprint = {gr-qc/0110100},
 primaryClass = {gr-qc},
       adsurl = {https://ui.adsabs.harvard.edu/abs/2002PhRvD..65h4011G}
}

@ARTICLE{2016ApJ...828...73K,
       author = {{Khan}, Fazeel Mahmood and {Fiacconi}, Davide and {Mayer}, Lucio and {Berczik}, Peter and {Just}, Andreas},
        title = "{Swift Coalescence of Supermassive Black Holes in Cosmological Mergers of Massive Galaxies}",
      journal = {\apj},
         year = 2016,
        month = sep,
       volume = {828},
       number = {2},
          eid = {73},
        pages = {73},
          doi = {10.3847/0004-637X/828/2/73},
archivePrefix = {arXiv},
       eprint = {1604.00015},
 primaryClass = {astro-ph.GA},
       adsurl = {https://ui.adsabs.harvard.edu/abs/2016ApJ...828...73K}
}

@ARTICLE{2021A&A...649A..41A,
       author = {{Avramov}, Branislav and {Berczik}, Peter and {Meiron}, Yohai and {Acharya}, Anshuman and {Just}, Andreas},
        title = "{Properties of loss cone stars in a cosmological galaxy merger remnant}",
      journal = {\aap},
         year = 2021,
        month = may,
       volume = {649},
          eid = {A41},
        pages = {A41},
          doi = {10.1051/0004-6361/202039698},
archivePrefix = {arXiv},
       eprint = {2011.08216},
 primaryClass = {astro-ph.GA},
       adsurl = {https://ui.adsabs.harvard.edu/abs/2021A&A...649A..41A}
}

@ARTICLE{2005PhRvD..72h4013H,
       author = {{Hawking}, S.~W.},
        title = "{Information loss in black holes}",
      journal = {\prd},
         year = 2005,
        month = oct,
       volume = {72},
       number = {8},
          eid = {084013},
        pages = {084013},
          doi = {10.1103/PhysRevD.72.084013},
archivePrefix = {arXiv},
       eprint = {hep-th/0507171},
 primaryClass = {hep-th},
       adsurl = {https://ui.adsabs.harvard.edu/abs/2005PhRvD..72h4013H}
}

@ARTICLE{2024A&A...685A..92S,
       author = {{Sala}, Luca and {Valentini}, Milena and {Biffi}, Veronica and {Dolag}, Klaus},
        title = "{Supermassive black hole spin evolution in cosmological simulations with OPENGADGET3}",
      journal = {\aap},
         year = 2024,
        month = may,
       volume = {685},
          eid = {A92},
        pages = {A92},
          doi = {10.1051/0004-6361/202348925},
archivePrefix = {arXiv},
       eprint = {2312.07657},
 primaryClass = {astro-ph.GA},
       adsurl = {https://ui.adsabs.harvard.edu/abs/2024A&A...685A..92S}
}

@ARTICLE{2017arXiv170307182W,
       author = {{Wang}, Yu and {Ghasemi-Nodehi}, M. and {Guainazzi}, Matteo and {Bambi}, Cosimo},
        title = "{The Spin of the Supermassive Black Hole in MCG-05-23-16}",
      journal = {arXiv e-prints},
         year = 2017,
        month = mar,
          eid = {arXiv:1703.07182},
        pages = {arXiv:1703.07182},
          doi = {10.48550/arXiv.1703.07182},
archivePrefix = {arXiv},
       eprint = {1703.07182},
 primaryClass = {astro-ph.HE},
       adsurl = {https://ui.adsabs.harvard.edu/abs/2017arXiv170307182W}
}

@article{10.1093/mnras/stz2836,
    author = {Bustamante, Sebastian and Springel, Volker},
    title = {Spin evolution and feedback of supermassive black holes in cosmological simulations},
    journal = {MNRAS},
    volume = {490},
    number = {3},
    pages = {4133-4153},
    year = {2019},
    month = {10},
    issn = {0035-8711},
    doi = {10.1093/mnras/stz2836},
    url = {https://doi.org/10.1093/mnras/stz2836},
    eprint = {https://academic.oup.com/mnras/article-pdf/490/3/4133/30380581/stz2836.pdf},
}

@ARTICLE{2018MNRAS.477.3807F,
       author = {{Fiacconi}, Davide and {Sijacki}, Debora and {Pringle}, J.~E.},
        title = "{Galactic nuclei evolution with spinning black holes: method and implementation}",
      journal = {\mnras},
         year = 2018,
        month = jul,
       volume = {477},
       number = {3},
        pages = {3807-3835},
          doi = {10.1093/mnras/sty893},
archivePrefix = {arXiv},
       eprint = {1712.00023},
 primaryClass = {astro-ph.GA},
       adsurl = {https://ui.adsabs.harvard.edu/abs/2018MNRAS.477.3807F}
}

@ARTICLE{2015ApJ...810..139H,
       author = {{Holley-Bockelmann}, Kelly and {Khan}, Fazeel Mahmood},
        title = "{Galaxy Rotation and Rapid Supermassive Binary Coalescence}",
      journal = {\apj},
         year = 2015,
        month = sep,
       volume = {810},
       number = {2},
          eid = {139},
        pages = {139},
          doi = {10.1088/0004-637X/810/2/139},
archivePrefix = {arXiv},
       eprint = {1505.06203},
 primaryClass = {astro-ph.GA},
       adsurl = {https://ui.adsabs.harvard.edu/abs/2015ApJ...810..139H}
}

@ARTICLE{2007NewA...12..357H,
       author = {{Harfst}, Stefan and {Gualandris}, Alessia and {Merritt}, David and {Spurzem}, Rainer and {Portegies Zwart}, Simon and {Berczik}, Peter},
        title = "{Performance analysis of direct N-body algorithms on special-purpose supercomputers}",
      journal = {\na},
         year = 2007,
        month = jul,
       volume = {12},
       number = {5},
        pages = {357-377},
          doi = {10.1016/j.newast.2006.11.003},
archivePrefix = {arXiv},
       eprint = {astro-ph/0608125},
 primaryClass = {astro-ph},
       adsurl = {https://ui.adsabs.harvard.edu/abs/2007NewA...12..357H}
}

@INPROCEEDINGS{2004cbhg.symp..263M,
       author = {{Merritt}, David},
        title = "{Single and Binary Black Holes and their Influence on Nuclear Structure}",
    booktitle = {Coevolution of Black Holes and Galaxies},
         year = 2004,
       editor = {{Ho}, Luis C.},
        month = jan,
        pages = {263},
          doi = {10.48550/arXiv.astro-ph/0301257},
archivePrefix = {arXiv},
       eprint = {astro-ph/0301257},
 primaryClass = {astro-ph},
       adsurl = {https://ui.adsabs.harvard.edu/abs/2004cbhg.symp..263M}
}

@ARTICLE{2017Natur.548..426F,
       author = {{Farr}, Will M. and {Stevenson}, Simon and {Miller}, M. Coleman and {Mandel}, Ilya and {Farr}, Ben and {Vecchio}, Alberto},
        title = "{Distinguishing spin-aligned and isotropic black hole populations with gravitational waves}",
      journal = {\nat},
         year = 2017,
        month = aug,
       volume = {548},
       number = {7667},
        pages = {426-429},
          doi = {10.1038/nature23453},
archivePrefix = {arXiv},
       eprint = {1706.01385},
 primaryClass = {astro-ph.HE},
       adsurl = {https://ui.adsabs.harvard.edu/abs/2017Natur.548..426F}
}

@ARTICLE{2015ARA&A..53..115K,
       author = {{King}, Andrew and {Pounds}, Ken},
        title = "{Powerful Outflows and Feedback from Active Galactic Nuclei}",
      journal = {\araa},
         year = 2015,
        month = aug,
       volume = {53},
        pages = {115-154},
          doi = {10.1146/annurev-astro-082214-122316},
archivePrefix = {arXiv},
       eprint = {1503.05206},
 primaryClass = {astro-ph.GA},
       adsurl = {https://ui.adsabs.harvard.edu/abs/2015ARA&A..53..115K}
}

@ARTICLE{2015MNRAS.446..521S,
       author = {{Schaye}, Joop and {Crain}, Robert A. and {Bower}, Richard G. and {Furlong}, Michelle and {Schaller}, Matthieu and {Theuns}, Tom and {Dalla Vecchia}, Claudio and {Frenk}, Carlos S. and {McCarthy}, I.~G. and {Helly}, John C. and {Jenkins}, Adrian and {Rosas-Guevara}, Y.~M. and {White}, Simon D.~M. and {Baes}, Maarten and {Booth}, C.~M. and {Camps}, Peter and {Navarro}, Julio F. and {Qu}, Yan and {Rahmati}, Alireza and {Sawala}, Till and {Thomas}, Peter A. and {Trayford}, James},
        title = "{The EAGLE project: simulating the evolution and assembly of galaxies and their environments}",
      journal = {\mnras},
         year = 2015,
        month = jan,
       volume = {446},
       number = {1},
        pages = {521-554},
          doi = {10.1093/mnras/stu2058},
archivePrefix = {arXiv},
       eprint = {1407.7040},
 primaryClass = {astro-ph.GA},
       adsurl = {https://ui.adsabs.harvard.edu/abs/2015MNRAS.446..521S}
}

@ARTICLE{2015MNRAS.451.1868T,
       author = {{Tremmel}, M. and {Governato}, F. and {Volonteri}, M. and {Quinn}, T.~R.},
        title = "{Off the beaten path: a new approach to realistically model the orbital decay of supermassive black holes in galaxy formation simulations}",
      journal = {\mnras},
         year = 2015,
        month = aug,
       volume = {451},
       number = {2},
        pages = {1868-1874},
          doi = {10.1093/mnras/stv1060},
archivePrefix = {arXiv},
       eprint = {1501.07609},
 primaryClass = {astro-ph.GA},
       adsurl = {https://ui.adsabs.harvard.edu/abs/2015MNRAS.451.1868T}
}

@ARTICLE{2009MNRAS.398...53B,
       author = {{Booth}, C.~M. and {Schaye}, Joop},
        title = "{Cosmological simulations of the growth of supermassive black holes and feedback from active galactic nuclei: method and tests}",
      journal = {\mnras},
         year = 2009,
        month = sep,
       volume = {398},
       number = {1},
        pages = {53-74},
          doi = {10.1111/j.1365-2966.2009.15043.x},
archivePrefix = {arXiv},
       eprint = {0904.2572},
 primaryClass = {astro-ph.CO},
       adsurl = {https://ui.adsabs.harvard.edu/abs/2009MNRAS.398...53B}
}

@ARTICLE{2008PhRvD..78d4021R,
       author = {{Racine}, {\'E}tienne},
        title = "{Analysis of spin precession in binary black hole systems including quadrupole-monopole interaction}",
      journal = {\prd},
         year = 2008,
        month = aug,
       volume = {78},
       number = {4},
          eid = {044021},
        pages = {044021},
          doi = {10.1103/PhysRevD.78.044021},
archivePrefix = {arXiv},
       eprint = {0803.1820},
 primaryClass = {gr-qc},
       adsurl = {https://ui.adsabs.harvard.edu/abs/2008PhRvD..78d4021R}
}

@ARTICLE{1995PhRvD..52..821K,
       author = {{Kidder}, Lawrence E.},
        title = "{Coalescing binary systems of compact objects to (post)$^{5/2}$-Newtonian order. V. Spin effects}",
      journal = {\prd},
         year = 1995,
        month = jul,
       volume = {52},
       number = {2},
        pages = {821-847},
          doi = {10.1103/PhysRevD.52.821},
archivePrefix = {arXiv},
       eprint = {gr-qc/9506022},
 primaryClass = {gr-qc},
       adsurl = {https://ui.adsabs.harvard.edu/abs/1995PhRvD..52..821K}
}

@ARTICLE{2017ApJ...837..135R,
       author = {{Rasskazov}, Alexander and {Merritt}, David},
        title = "{Evolution of Binary Supermassive Black Holes in Rotating Nuclei}",
      journal = {\apj},
         year = 2017,
        month = mar,
       volume = {837},
       number = {2},
          eid = {135},
        pages = {135},
          doi = {10.3847/1538-4357/aa6188},
archivePrefix = {arXiv},
       eprint = {1610.08555},
 primaryClass = {astro-ph.GA},
       adsurl = {https://ui.adsabs.harvard.edu/abs/2017ApJ...837..135R}
}

@ARTICLE{2015ApJ...810...49V,
       author = {{Vasiliev}, Eugene and {Antonini}, Fabio and {Merritt}, David},
        title = "{The Final-parsec Problem in the Collisionless Limit}",
      journal = {\apj},
         year = 2015,
        month = sep,
       volume = {810},
       number = {1},
          eid = {49},
        pages = {49},
          doi = {10.1088/0004-637X/810/1/49},
archivePrefix = {arXiv},
       eprint = {1505.05480},
 primaryClass = {astro-ph.GA},
       adsurl = {https://ui.adsabs.harvard.edu/abs/2015ApJ...810...49V}
}

@ARTICLE{2014ApJ...785..163V,
       author = {{Vasiliev}, Eugene and {Antonini}, Fabio and {Merritt}, David},
        title = "{The Final-parsec Problem in Nonspherical Galaxies Revisited}",
      journal = {\apj},
         year = 2014,
        month = apr,
       volume = {785},
       number = {2},
          eid = {163},
        pages = {163},
          doi = {10.1088/0004-637X/785/2/163},
archivePrefix = {arXiv},
       eprint = {1311.1167},
 primaryClass = {astro-ph.GA},
       adsurl = {https://ui.adsabs.harvard.edu/abs/2014ApJ...785..163V}
}

@ARTICLE{2010ApJ...719..851S,
       author = {{Sesana}, Alberto},
        title = "{Self Consistent Model for the Evolution of Eccentric Massive Black Hole Binaries in Stellar Environments: Implications for Gravitational Wave Observations}",
      journal = {\apj},
         year = 2010,
        month = aug,
       volume = {719},
       number = {1},
        pages = {851-864},
          doi = {10.1088/0004-637X/719/1/851},
archivePrefix = {arXiv},
       eprint = {1006.0730},
 primaryClass = {astro-ph.CO},
       adsurl = {https://ui.adsabs.harvard.edu/abs/2010ApJ...719..851S}
}

@ARTICLE{2017MNRAS.464.2301G,
       author = {{Gualandris}, Alessia and {Read}, Justin I. and {Dehnen}, Walter and {Bortolas}, Elisa},
        title = "{Collisionless loss-cone refilling: there is no final parsec problem}",
      journal = {\mnras},
         year = 2017,
        month = jan,
       volume = {464},
       number = {2},
        pages = {2301-2310},
          doi = {10.1093/mnras/stw2528},
archivePrefix = {arXiv},
       eprint = {1609.09383},
 primaryClass = {astro-ph.GA},
       adsurl = {https://ui.adsabs.harvard.edu/abs/2017MNRAS.464.2301G}
}

@ARTICLE{2013ApJ...773..100K,
       author = {{Khan}, Fazeel Mahmood and {Holley-Bockelmann}, Kelly and {Berczik}, Peter and {Just}, Andreas},
        title = "{Supermassive Black Hole Binary Evolution in Axisymmetric Galaxies: The Final Parsec Problem is Not a Problem}",
      journal = {\apj},
         year = 2013,
        month = aug,
       volume = {773},
       number = {2},
          eid = {100},
        pages = {100},
          doi = {10.1088/0004-637X/773/2/100},
archivePrefix = {arXiv},
       eprint = {1302.1871},
 primaryClass = {astro-ph.GA},
       adsurl = {https://ui.adsabs.harvard.edu/abs/2013ApJ...773..100K}
}

@ARTICLE{2000ApJ...539L..13G,
       author = {{Gebhardt}, Karl and {Bender}, Ralf and {Bower}, Gary and {Dressler}, Alan and {Faber}, S.~M. and {Filippenko}, Alexei V. and {Green}, Richard and {Grillmair}, Carl and {Ho}, Luis C. and {Kormendy}, John and {Lauer}, Tod R. and {Magorrian}, John and {Pinkney}, Jason and {Richstone}, Douglas and {Tremaine}, Scott},
        title = "{A Relationship between Nuclear Black Hole Mass and Galaxy Velocity Dispersion}",
      journal = {\apjl},
         year = 2000,
        month = aug,
       volume = {539},
       number = {1},
        pages = {L13-L16},
          doi = {10.1086/312840},
archivePrefix = {arXiv},
       eprint = {astro-ph/0006289},
 primaryClass = {astro-ph},
       adsurl = {https://ui.adsabs.harvard.edu/abs/2000ApJ...539L..13G}
}

@ARTICLE{2011ApJ...728...12L,
       author = {{Lo}, Ka-Wai and {Lin}, Lap-Ming},
        title = "{The Spin Parameter of Uniformly Rotating Compact Stars}",
      journal = {\apj},
         year = 2011,
        month = feb,
       volume = {728},
       number = {1},
          eid = {12},
        pages = {12},
          doi = {10.1088/0004-637X/728/1/12},
archivePrefix = {arXiv},
       eprint = {1011.3563},
 primaryClass = {astro-ph.HE},
       adsurl = {https://ui.adsabs.harvard.edu/abs/2011ApJ...728...12L}
}

@ARTICLE{2016PhRvL.116x1102A,
       author = {{Abbott}, B.~P. and {Abbott}, R. and {Abbott}, T.~D. and {Abernathy}, M.~R. and {Acernese}, F. and {Ackley}, K. and {Adams}, C. and {Adams}, T. and {Addesso}, P. and {Adhikari}, R.~X. and {Adya}, V.~B. and {Affeldt}, C. and {Agathos}, M. and {Agatsuma}, K. and {Aggarwal}, N. and {Aguiar}, O.~D. and {Aiello}, L. and {Ain}, A. and {Ajith}, P. and {Allen}, B. and {Allocca}, A. and {Altin}, P.~A. and {Anderson}, S.~B. and {Anderson}, W.~G. and {Arai}, K. and {Araya}, M.~C. and {Arceneaux}, C.~C. and {Areeda}, J.~S. and {Arnaud}, N. and {Arun}, K.~G. and {Ascenzi}, S. and {Ashton}, G. and {Ast}, M. and {Aston}, S.~M. and {Astone}, P. and {Aufmuth}, P. and {Aulbert}, C. and {Babak}, S. and {Bacon}, P. and {Bader}, M.~K.~M. and {Baker}, P.~T. and {Baldaccini}, F. and {Ballardin}, G. and {Ballmer}, S.~W. and {Barayoga}, J.~C. and {Barclay}, S.~E. and {Barish}, B.~C. and {Barker}, D. and {Barone}, F. and {Barr}, B. and {Barsotti}, L. and {Barsuglia}, M. and {Barta}, D. and {Bartlett}, J. and {Bartos}, I. and {Bassiri}, R. and {Basti}, A. and {Batch}, J.~C. and {Baune}, C. and {Bavigadda}, V. and {Bazzan}, M. and {Behnke}, B. and {Bejger}, M. and {Bell}, A.~S. and {Bell}, C.~J. and {Berger}, B.~K. and {Bergman}, J. and {Bergmann}, G. and {Berry}, C.~P.~L. and {Bersanetti}, D. and {Bertolini}, A. and {Betzwieser}, J. and {Bhagwat}, S. and {Bhandare}, R. and {Bilenko}, I.~A. and {Billingsley}, G. and {Birch}, J. and {Birney}, I.~A. and {Birnholtz}, O. and {Biscans}, S. and {Bisht}, A. and {Bitossi}, M. and {Biwer}, C. and {Bizouard}, M.~A. and {Blackburn}, J.~K. and {Blair}, C.~D. and {Blair}, D.~G. and {Blair}, R.~M. and {Bloemen}, S. and {Bock}, O. and {Bodiya}, T.~P. and {Boer}, M. and {Bogaert}, G. and {Bogan}, C. and {Bohe}, A. and {Bojtos}, P. and {Bond}, C. and {Bondu}, F. and {Bonnand}, R. and {Boom}, B.~A. and {Bork}, R. and {Boschi}, V. and {Bose}, S. and {Bouffanais}, Y. and {Bozzi}, A. and {Bradaschia}, C. and {Brady}, P.~R. and {Braginsky}, V.~B. and {Branchesi}, M. and {Brau}, J.~E. and {Briant}, T. and {Brillet}, A. and {Brinkmann}, M. and {Brisson}, V. and {Brockill}, P. and {Brooks}, A.~F. and {Brown}, D.~A. and {Brown}, D.~D. and {Brown}, N.~M. and {Buchanan}, C.~C. and {Buikema}, A. and {Bulik}, T. and {Bulten}, H.~J. and {Buonanno}, A. and {Buskulic}, D. and {Buy}, C. and {Byer}, R.~L. and {Cadonati}, L. and {Cagnoli}, G. and {Cahillane}, C. and {Calder{\'o}n Bustillo}, J. and {Callister}, T. and {Calloni}, E. and {Camp}, J.~B. and {Cannon}, K.~C. and {Cao}, J. and {Capano}, C.~D. and {Capocasa}, E. and {Carbognani}, F. and {Caride}, S. and {Casanueva Diaz}, J. and {Casentini}, C. and {Caudill}, S. and {Cavagli{\`a}}, M. and {Cavalier}, F. and {Cavalieri}, R. and {Cella}, G. and {Cepeda}, C.~B. and {Cerboni Baiardi}, L. and {Cerretani}, G. and {Cesarini}, E. and {Chakraborty}, R. and {Chalermsongsak}, T. and {Chamberlin}, S.~J. and {Chan}, M. and {Chao}, S. and {Charlton}, P. and {Chassande-Mottin}, E. and {Chen}, H.~Y. and {Chen}, Y. and {Cheng}, C. and {Chincarini}, A. and {Chiummo}, A. and {Cho}, H.~S. and {Cho}, M. and {Chow}, J.~H. and {Christensen}, N. and {Chu}, Q. and {Chua}, S. and {Chung}, S. and {Ciani}, G. and {Clara}, F. and {Clark}, J.~A. and {Cleva}, F. and {Coccia}, E. and {Cohadon}, P.-F. and {Colla}, A. and {Collette}, C.~G. and {Cominsky}, L. and {Constancio}, M. and {Conte}, A. and {Conti}, L. and {Cook}, D. and {Corbitt}, T.~R. and {Cornish}, N. and {Corsi}, A. and {Cortese}, S. and {Costa}, C.~A. and {Coughlin}, M.~W. and {Coughlin}, S.~B. and {Coulon}, J.-P. and {Countryman}, S.~T. and {Couvares}, P. and {Cowan}, E.~E. and {Coward}, D.~M. and {Cowart}, M.~J. and {Coyne}, D.~C. and {Coyne}, R. and {Craig}, K. and {Creighton}, J.~D.~E.},
        title = "{Properties of the Binary Black Hole Merger GW150914}",
      journal = {\prl},
         year = 2016,
        month = jun,
       volume = {116},
       number = {24},
          eid = {241102},
        pages = {241102},
          doi = {10.1103/PhysRevLett.116.241102},
archivePrefix = {arXiv},
       eprint = {1602.03840},
 primaryClass = {gr-qc},
       adsurl = {https://ui.adsabs.harvard.edu/abs/2016PhRvL.116x1102A}
}

@ARTICLE{2011PhRvL.106x1101A,
       author = {{Ajith}, P. and {Hannam}, M. and {Husa}, S. and {Chen}, Y. and {Br{\"u}gmann}, B. and {Dorband}, N. and {M{\"u}ller}, D. and {Ohme}, F. and {Pollney}, D. and {Reisswig}, C. and {Santamar{\'\i}a}, L. and {Seiler}, J.},
        title = "{Inspiral-Merger-Ringdown Waveforms for Black-Hole Binaries with Nonprecessing Spins}",
      journal = {\prl},
         year = 2011,
        month = jun,
       volume = {106},
       number = {24},
          eid = {241101},
        pages = {241101},
          doi = {10.1103/PhysRevLett.106.241101},
archivePrefix = {arXiv},
       eprint = {0909.2867},
 primaryClass = {gr-qc},
       adsurl = {https://ui.adsabs.harvard.edu/abs/2011PhRvL.106x1101A}
}

@ARTICLE{2009ApJ...704L..40B,
       author = {{Barausse}, Enrico and {Rezzolla}, Luciano},
        title = "{Predicting the Direction of the Final Spin from the Coalescence of Two Black Holes}",
      journal = {\apjl},
         year = 2009,
        month = oct,
       volume = {704},
       number = {1},
        pages = {L40-L44},
          doi = {10.1088/0004-637X/704/1/L40},
archivePrefix = {arXiv},
       eprint = {0904.2577},
 primaryClass = {gr-qc},
       adsurl = {https://ui.adsabs.harvard.edu/abs/2009ApJ...704L..40B}
}

@ARTICLE{2018PhRvD..98h4036G,
       author = {{Gerosa}, Davide and {Berti}, Emanuele and {O'Shaughnessy}, Richard and {Belczynski}, Krzysztof and {Kesden}, Michael and {Wysocki}, Daniel and {Gladysz}, Wojciech},
        title = "{Spin orientations of merging black holes formed from the evolution of stellar binaries}",
      journal = {\prd},
         year = 2018,
        month = oct,
       volume = {98},
       number = {8},
          eid = {084036},
        pages = {084036},
          doi = {10.1103/PhysRevD.98.084036},
archivePrefix = {arXiv},
       eprint = {1808.02491},
 primaryClass = {astro-ph.HE},
       adsurl = {https://ui.adsabs.harvard.edu/abs/2018PhRvD..98h4036G}
}

@ARTICLE{2019ApJ...882L..24A,
       author = {{Abbott}, B.~P. and {Abbott}, R. and {Abbott}, T.~D. and {Abraham}, S. and {Acernese}, F. and {Ackley}, K. and {Adams}, C. and {Adhikari}, R.~X. and {Adya}, V.~B. and {Affeldt}, C. and {Agathos}, M. and {Agatsuma}, K. and {Aggarwal}, N. and {Aguiar}, O.~D. and {Aiello}, L. and {Ain}, A. and {Ajith}, P. and {Allen}, G. and {Allocca}, A. and {Aloy}, M.~A. and {Altin}, P.~A. and {Amato}, A. and {Ananyeva}, A. and {Anderson}, S.~B. and {Anderson}, W.~G. and {Angelova}, S.~V. and {Antier}, S. and {Appert}, S. and {Arai}, K. and {Araya}, M.~C. and {Areeda}, J.~S. and {Ar{\`e}ne}, M. and {Arnaud}, N. and {Arun}, K.~G. and {Ascenzi}, S. and {Ashton}, G. and {Aston}, S.~M. and {Astone}, P. and {Aubin}, F. and {Aufmuth}, P. and {AultONeal}, K. and {Austin}, C. and {Avendano}, V. and {Avila-Alvarez}, A. and {Babak}, S. and {Bacon}, P. and {Badaracco}, F. and {Bader}, M.~K.~M. and {Bae}, S. and {Baker}, P.~T. and {Baldaccini}, F. and {Ballardin}, G. and {Ballmer}, S.~W. and {Banagiri}, S. and {Barayoga}, J.~C. and {Barclay}, S.~E. and {Barish}, B.~C. and {Barker}, D. and {Barkett}, K. and {Barnum}, S. and {Barone}, F. and {Barr}, B. and {Barsotti}, L. and {Barsuglia}, M. and {Barta}, D. and {Bartlett}, J. and {Bartos}, I. and {Bassiri}, R. and {Basti}, A. and {Bawaj}, M. and {Bayley}, J.~C. and {Bazzan}, M. and {B{\'e}csy}, B. and {Bejger}, M. and {Belahcene}, I. and {Bell}, A.~S. and {Beniwal}, D. and {Berger}, B.~K. and {Bergmann}, G. and {Bernuzzi}, S. and {Bero}, J.~J. and {Berry}, C.~P.~L. and {Bersanetti}, D. and {Bertolini}, A. and {Betzwieser}, J. and {Bhandare}, R. and {Bidler}, J. and {Bilenko}, I.~A. and {Bilgili}, S.~A. and {Billingsley}, G. and {Birch}, J. and {Birney}, R. and {Birnholtz}, O. and {Biscans}, S. and {Biscoveanu}, S. and {Bisht}, A. and {Bitossi}, M. and {Bizouard}, M.~A. and {Blackburn}, J.~K. and {Blair}, C.~D. and {Blair}, D.~G. and {Blair}, R.~M. and {Bloemen}, S. and {Bode}, N. and {Boer}, M. and {Boetzel}, Y. and {Bogaert}, G. and {Bondu}, F. and {Bonilla}, E. and {Bonnand}, R. and {Booker}, P. and {Boom}, B.~A. and {Booth}, C.~D. and {Bork}, R. and {Boschi}, V. and {Bose}, S. and {Bossie}, K. and {Bossilkov}, V. and {Bosveld}, J. and {Bouffanais}, Y. and {Bozzi}, A. and {Bradaschia}, C. and {Brady}, P.~R. and {Bramley}, A. and {Branchesi}, M. and {Brau}, J.~E. and {Briant}, T. and {Briggs}, J.~H. and {Brighenti}, F. and {Brillet}, A. and {Brinkmann}, M. and {Brisson}, V. and {Brockill}, P. and {Brooks}, A.~F. and {Brown}, D.~D. and {Brunett}, S. and {Buikema}, A. and {Bulik}, T. and {Bulten}, H.~J. and {Buonanno}, A. and {Buscicchio}, R. and {Buskulic}, D. and {Buy}, C. and {Byer}, R.~L. and {Cabero}, M. and {Cadonati}, L. and {Cagnoli}, G. and {Cahillane}, C. and {Calder{\'o}n Bustillo}, J. and {Callister}, T.~A. and {Calloni}, E. and {Camp}, J.~B. and {Campbell}, W.~A. and {Canepa}, M. and {Cannon}, K.~C. and {Cao}, H. and {Cao}, J. and {Capocasa}, E. and {Carbognani}, F. and {Caride}, S. and {Carney}, M.~F. and {Carullo}, G. and {Casanueva Diaz}, J. and {Casentini}, C. and {Caudill}, S. and {Cavagli{\`a}}, M. and {Cavalier}, F. and {Cavalieri}, R. and {Cella}, G. and {Cerd{\'a}-Dur{\'a}n}, P. and {Cerretani}, G. and {Cesarini}, E. and {Chaibi}, O. and {Chakravarti}, K. and {Chamberlin}, S.~J. and {Chan}, M. and {Chao}, S. and {Charlton}, P. and {Chase}, E.~A. and {Chassande-Mottin}, E. and {Chatterjee}, D. and {Chaturvedi}, M. and {Chatziioannou}, K. and {Cheeseboro}, B.~D. and {Chen}, H.~Y. and {Chen}, X. and {Chen}, Y. and {Cheng}, H.-P. and {Cheong}, C.~K. and {Chia}, H.~Y. and {Chincarini}, A. and {Chiummo}, A. and {Cho}, G. and {Cho}, H.~S. and {Cho}, M. and {Christensen}, N. and {Chu}, Q. and {Chua}, S. and {Chung}, K.~W. and {Chung}, S.},
        title = "{Binary Black Hole Population Properties Inferred from the First and Second Observing Runs of Advanced LIGO and Advanced Virgo}",
      journal = {\apjl},
         year = 2019,
        month = sep,
       volume = {882},
       number = {2},
          eid = {L24},
        pages = {L24},
          doi = {10.3847/2041-8213/ab3800},
archivePrefix = {arXiv},
       eprint = {1811.12940},
 primaryClass = {astro-ph.HE},
       adsurl = {https://ui.adsabs.harvard.edu/abs/2019ApJ...882L..24A}
}

@ARTICLE{2014PhRvL.113o1101H,
       author = {{Hannam}, Mark and {Schmidt}, Patricia and {Boh{\'e}}, Alejandro and {Haegel}, Le{\"\i}la and {Husa}, Sascha and {Ohme}, Frank and {Pratten}, Geraint and {P{\"u}rrer}, Michael},
        title = "{Simple Model of Complete Precessing Black-Hole-Binary Gravitational Waveforms}",
      journal = {\prl},
         year = 2014,
        month = oct,
       volume = {113},
       number = {15},
          eid = {151101},
        pages = {151101},
          doi = {10.1103/PhysRevLett.113.151101},
archivePrefix = {arXiv},
       eprint = {1308.3271},
 primaryClass = {gr-qc},
       adsurl = {https://ui.adsabs.harvard.edu/abs/2014PhRvL.113o1101H}
}

@ARTICLE{2020A&A...635A..97B,
       author = {{Bavera}, Simone S. and {Fragos}, Tassos and {Qin}, Ying and {Zapartas}, Emmanouil and {Neijssel}, Coenraad J. and {Mandel}, Ilya and {Batta}, Aldo and {Gaebel}, Sebastian M. and {Kimball}, Chase and {Stevenson}, Simon},
        title = "{The origin of spin in binary black holes. Predicting the distributions of the main observables of Advanced LIGO}",
      journal = {\aap},
         year = 2020,
        month = mar,
       volume = {635},
          eid = {A97},
        pages = {A97},
          doi = {10.1051/0004-6361/201936204},
archivePrefix = {arXiv},
       eprint = {1906.12257},
 primaryClass = {astro-ph.HE},
       adsurl = {https://ui.adsabs.harvard.edu/abs/2020A&A...635A..97B}
}

@ARTICLE{2000AnP...512..378J,
       author = {{Jaranowski}, P. and {Sch{\"a}fer}, G.},
        title = "{The binary black-hole dynamics at the third post-Newtonian order in the orbital motion}",
      journal = {Annalen der Physik},
         year = 2000,
        month = may,
       volume = {512},
       number = {3},
        pages = {378-383},
          doi = {10.1002/(SICI)1521-3889(200005)9:3/5<378::AID-ANDP378>3.0.CO;2-M},
archivePrefix = {arXiv},
       eprint = {gr-qc/0003054},
 primaryClass = {gr-qc},
       adsurl = {https://ui.adsabs.harvard.edu/abs/2000AnP...512..378J}
}

@ARTICLE{2006PhRvD..74j4033F,
       author = {{Faye}, Guillaume and {Blanchet}, Luc and {Buonanno}, Alessandra},
        title = "{Higher-order spin effects in the dynamics of compact binaries. I. Equations of motion}",
      journal = {\prd},
         year = 2006,
        month = nov,
       volume = {74},
       number = {10},
          eid = {104033},
        pages = {104033},
          doi = {10.1103/PhysRevD.74.104033},
archivePrefix = {arXiv},
       eprint = {gr-qc/0605139},
 primaryClass = {gr-qc},
       adsurl = {https://ui.adsabs.harvard.edu/abs/2006PhRvD..74j4033F}
}

@ARTICLE{1994PhRvD..49.6274A,
       author = {{Apostolatos}, Theocharis A. and {Cutler}, Curt and {Sussman}, Gerald J. and {Thorne}, Kip S.},
        title = "{Spin-induced orbital precession and its modulation of the gravitational waveforms from merging binaries}",
      journal = {\prd},
         year = 1994,
        month = jun,
       volume = {49},
       number = {12},
        pages = {6274-6297},
          doi = {10.1103/PhysRevD.49.6274},
       adsurl = {https://ui.adsabs.harvard.edu/abs/1994PhRvD..49.6274A}
}

@ARTICLE{2001PhRvD..63d4006T,
       author = {{Tagoshi}, Hideyuki and {Ohashi}, Akira and {Owen}, Benjamin J.},
        title = "{Gravitational field and equations of motion of spinning compact binaries to 2.5 post-Newtonian order}",
      journal = {\prd},
         year = 2001,
        month = feb,
       volume = {63},
       number = {4},
          eid = {044006},
        pages = {044006},
          doi = {10.1103/PhysRevD.63.044006},
archivePrefix = {arXiv},
       eprint = {gr-qc/0010014},
 primaryClass = {gr-qc},
       adsurl = {https://ui.adsabs.harvard.edu/abs/2001PhRvD..63d4006T}
}

@ARTICLE{1979GReGr..11..149B,
       author = {{Barker}, B.~M. and {O'Connell}, R.~F.},
        title = "{The gravitational interaction: spin, rotation, and quantum effects - a review.}",
      journal = {General Relativity and Gravitation},
         year = 1979,
        month = oct,
       volume = {11},
       number = {2},
        pages = {149-175},
          doi = {10.1007/BF00756587},
       adsurl = {https://ui.adsabs.harvard.edu/abs/1979GReGr..11..149B}
}

@ARTICLE{1975PhRvD..12..329B,
       author = {{Barker}, B.~M. and {O'Connell}, R.~F.},
        title = "{Gravitational two-body problem with arbitrary masses, spins, and quadrupole moments}",
      journal = {\prd},
         year = 1975,
        month = jul,
       volume = {12},
       number = {2},
        pages = {329-335},
          doi = {10.1103/PhysRevD.12.329},
       adsurl = {https://ui.adsabs.harvard.edu/abs/1975PhRvD..12..329B}
}

@ARTICLE{2007PhRvL..98w1101G,
       author = {{Gonz{\'a}lez}, Jos{\'e} A. and {Hannam}, Mark and {Sperhake}, Ulrich and {Br{\"u}gmann}, Bernd and {Husa}, Sascha},
        title = "{Supermassive Recoil Velocities for Binary Black-Hole Mergers with Antialigned Spins}",
      journal = {\prl},
         year = 2007,
        month = jun,
       volume = {98},
       number = {23},
          eid = {231101},
        pages = {231101},
          doi = {10.1103/PhysRevLett.98.231101},
archivePrefix = {arXiv},
       eprint = {gr-qc/0702052},
 primaryClass = {gr-qc},
       adsurl = {https://ui.adsabs.harvard.edu/abs/2007PhRvL..98w1101G}
}

@ARTICLE{2001PhRvD..64l4013D,
       author = {{Damour}, Thibault},
        title = "{Coalescence of two spinning black holes: An effective one-body approach}",
      journal = {\prd},
         year = 2001,
        month = dec,
       volume = {64},
       number = {12},
        pages = {124013},
          doi = {10.1103/PhysRevD.64.124013},
archivePrefix = {arXiv},
       eprint = {gr-qc/0103018},
 primaryClass = {gr-qc},
       adsurl = {https://ui.adsabs.harvard.edu/abs/2001PhRvD..64l4013D}
}

@ARTICLE{1999PhRvD..59h4006B,
       author = {{Buonanno}, A. and {Damour}, T.},
        title = "{Effective one-body approach to general relativistic two-body dynamics}",
      journal = {\prd},
         year = 1999,
        month = apr,
       volume = {59},
       number = {8},
          eid = {084006},
        pages = {084006},
          doi = {10.1103/PhysRevD.59.084006},
archivePrefix = {arXiv},
       eprint = {gr-qc/9811091},
 primaryClass = {gr-qc},
       adsurl = {https://ui.adsabs.harvard.edu/abs/1999PhRvD..59h4006B}
}

@ARTICLE{2018LRR....21....7S,
       author = {{Sch{\"a}fer}, Gerhard and {Jaranowski}, Piotr},
        title = "{Hamiltonian formulation of general relativity and post-Newtonian dynamics of compact binaries}",
      journal = {Living Reviews in Relativity},
         year = 2018,
        month = dec,
       volume = {21},
       number = {1},
          eid = {7},
        pages = {7},
          doi = {10.1007/s41114-018-0016-5},
archivePrefix = {arXiv},
       eprint = {1805.07240},
 primaryClass = {gr-qc},
       adsurl = {https://ui.adsabs.harvard.edu/abs/2018LRR....21....7S}
}

@ARTICLE{2018PhRvD..97f4010V,
       author = {{Vines}, Justin and {Steinhoff}, Jan},
        title = "{Spin-multipole effects in binary black holes and the test-body limit}",
      journal = {\prd},
         year = 2018,
        month = mar,
       volume = {97},
       number = {6},
          eid = {064010},
        pages = {064010},
          doi = {10.1103/PhysRevD.97.064010},
archivePrefix = {arXiv},
       eprint = {1606.08832},
 primaryClass = {gr-qc},
       adsurl = {https://ui.adsabs.harvard.edu/abs/2018PhRvD..97f4010V}
}

@ARTICLE{1993PhRvD..47.4183K,
       author = {{Kidder}, Lawrence E. and {Will}, Clifford M. and {Wiseman}, Alan G.},
        title = "{Spin effects in the inspiral of coalescing compact binaries}",
      journal = {\prd},
         year = 1993,
        month = may,
       volume = {47},
       number = {10},
        pages = {R4183-R4187},
          doi = {10.1103/PhysRevD.47.R4183},
archivePrefix = {arXiv},
       eprint = {gr-qc/9211025},
 primaryClass = {gr-qc},
       adsurl = {https://ui.adsabs.harvard.edu/abs/1993PhRvD..47.4183K}
}

@ARTICLE{1992PASJ...44..141M,
       author = {{Makino}, Junichiro and {Aarseth}, Sverre J.},
        title = "{On a Hermite Integrator with Ahmad-Cohen Scheme for Gravitational Many-Body Problems}",
      journal = {\pasj},
         year = 1992,
        month = may,
       volume = {44},
       number = {2},
        pages = {141-151},
          doi = {10.1093/pasj/44.2.141},
       adsurl = {https://ui.adsabs.harvard.edu/abs/1992PASJ...44..141M}
}

@ARTICLE{2006PhRvD..74j4005B,
       author = {{Buonanno}, Alessandra and {Chen}, Yanbei and {Damour}, Thibault},
        title = "{Transition from inspiral to plunge in precessing binaries of spinning black holes}",
      journal = {\prd},
         year = 2006,
        month = nov,
       volume = {74},
       number = {10},
          eid = {104005},
        pages = {104005},
          doi = {10.1103/PhysRevD.74.104005},
archivePrefix = {arXiv},
       eprint = {gr-qc/0508067},
 primaryClass = {gr-qc},
       adsurl = {https://ui.adsabs.harvard.edu/abs/2006PhRvD..74j4005B}
}

@article{PhysRevD.103.124026,
  title = {A taxonomy of black-hole binary spin precession and nutation},
  author = {Gangardt, Daria and Steinle, Nathan and Kesden, Michael and Gerosa, Davide and Stoikos, Evangelos},
  journal = {Phys. Rev. D},
  volume = {103},
  issue = {12},
  pages = {124026},
  numpages = {14},
  year = {2021},
  month = {Jun},
  publisher = {American Physical Society},
  doi = {10.1103/PhysRevD.103.124026},
  url = {https://link.aps.org/doi/10.1103/PhysRevD.103.124026}
}

@ARTICLE{2012ApJ...749..147K,
       author = {{Khan}, Fazeel Mahmood and {Preto}, Miguel and {Berczik}, Peter and {Berentzen}, Ingo and {Just}, Andreas and {Spurzem}, Rainer},
        title = "{Mergers of Unequal-mass Galaxies: Supermassive Black Hole Binary Evolution and Structure of Merger Remnants}",
      journal = {\apj},
         year = 2012,
        month = apr,
       volume = {749},
       number = {2},
          eid = {147},
        pages = {147},
          doi = {10.1088/0004-637X/749/2/147},
archivePrefix = {arXiv},
       eprint = {1202.2124},
 primaryClass = {astro-ph.CO},
       adsurl = {https://ui.adsabs.harvard.edu/abs/2012ApJ...749..147K}
}

@ARTICLE{2015ApJ...798..103K,
       author = {{Khan}, Fazeel Mahmood and {Holley-Bockelmann}, Kelly and {Berczik}, Peter},
        title = "{Ultramassive Black Hole Coalescence}",
      journal = {\apj},
         year = 2015,
        month = jan,
       volume = {798},
       number = {2},
          eid = {103},
        pages = {103},
          doi = {10.1088/0004-637X/798/2/103},
archivePrefix = {arXiv},
       eprint = {1405.6425},
 primaryClass = {astro-ph.GA},
       adsurl = {https://ui.adsabs.harvard.edu/abs/2015ApJ...798..103K}
}

@ARTICLE{2013CQGra..30x4004R,
       author = {{Reynolds}, Christopher S.},
        title = "{The spin of supermassive black holes}",
      journal = {Classical and Quantum Gravity},
         year = 2013,
        month = dec,
       volume = {30},
       number = {24},
          eid = {244004},
        pages = {244004},
          doi = {10.1088/0264-9381/30/24/244004},
archivePrefix = {arXiv},
       eprint = {1307.3246},
 primaryClass = {astro-ph.HE},
       adsurl = {https://ui.adsabs.harvard.edu/abs/2013CQGra..30x4004R}
}

@ARTICLE{2021ARA&A..59..117R,
       author = {{Reynolds}, Christopher S.},
        title = "{Observational Constraints on Black Hole Spin}",
      journal = {\araa},
         year = 2021,
        month = sep,
       volume = {59},
        pages = {117-154},
          doi = {10.1146/annurev-astro-112420-035022},
archivePrefix = {arXiv},
       eprint = {2011.08948},
 primaryClass = {astro-ph.HE},
       adsurl = {https://ui.adsabs.harvard.edu/abs/2021ARA&A..59..117R}
}

@BOOK{2013mams.book.....B,
       author = {{Brenneman}, Laura},
        title = "{Measuring the Angular Momentum of Supermassive Black Holes}",
         year = 2013,
         publisher = {Springer New York, NY},
          doi = {10.1007/978-1-4614-7771-6},
       adsurl = {https://ui.adsabs.harvard.edu/abs/2013mams.book.....B}
}

@ARTICLE{2021SSRv..217...65B,
       author = {{Bambi}, Cosimo and {Brenneman}, Laura W. and {Dauser}, Thomas and {Garc{\'\i}a}, Javier A. and {Grinberg}, Victoria and {Ingram}, Adam and {Jiang}, Jiachen and {Liu}, Honghui and {Lohfink}, Anne M. and {Marinucci}, Andrea and {Mastroserio}, Guglielmo and {Middei}, Riccardo and {Nampalliwar}, Sourabh and {Nied{\'z}wiecki}, Andrzej and {Steiner}, James F. and {Tripathi}, Ashutosh and {Zdziarski}, Andrzej A.},
        title = "{Towards Precision Measurements of Accreting Black Holes Using X-Ray Reflection Spectroscopy}",
      journal = {\ssr},
         year = 2021,
        month = aug,
       volume = {217},
       number = {5},
          eid = {65},
        pages = {65},
          doi = {10.1007/s11214-021-00841-8},
archivePrefix = {arXiv},
       eprint = {2011.04792},
 primaryClass = {astro-ph.HE},
       adsurl = {https://ui.adsabs.harvard.edu/abs/2021SSRv..217...65B}
}

@ARTICLE{2025ApJ...980..136R,
       author = {{Ricarte}, Angelo and {Natarajan}, Priyamvada and {Narayan}, Ramesh and {Palumbo}, Daniel C.~M.},
        title = "{Multimessenger Probes of Supermassive Black Hole Spin Evolution}",
      journal = {\apj},
         year = 2025,
        month = feb,
       volume = {980},
       number = {1},
          eid = {136},
        pages = {136},
          doi = {10.3847/1538-4357/ad9ea9},
archivePrefix = {arXiv},
       eprint = {2410.07477},
 primaryClass = {astro-ph.HE},
       adsurl = {https://ui.adsabs.harvard.edu/abs/2025ApJ...980..136R}
}

@ARTICLE{2024MNRAS.52710867B,
       author = {{Beckmann}, R.~S. and {Smethurst}, R.~J. and {Simmons}, B.~D. and {Coil}, A. and {Dubois}, Y. and {Garland}, I.~L. and {Lintott}, C.~J. and {Martin}, G. and {Peirani}, S. and {Pichon}, C.},
        title = "{Supermassive black holes in merger-free galaxies have higher spins which are preferentially aligned with their host galaxy}",
      journal = {\mnras},
         year = 2024,
        month = feb,
       volume = {527},
       number = {4},
        pages = {10867-10877},
          doi = {10.1093/mnras/stad1795},
       adsurl = {https://ui.adsabs.harvard.edu/abs/2024MNRAS.52710867B}
}

@ARTICLE{2025MNRAS.536.1838B,
       author = {{Beckmann}, R.~S. and {Dubois}, Y. and {Volonteri}, M. and {Dong-Paez}, C.~A. and {Peirani}, S. and {Piotrowska}, J.~M. and {Martin}, G. and {Kraljic}, K. and {Devriendt}, J. and {Pichon}, C. and {Yi}, S.~K.},
        title = "{Black hole spin evolution across cosmic time from the NEWHORIZON simulation}",
      journal = {\mnras},
         year = 2025,
        month = jan,
       volume = {536},
       number = {2},
        pages = {1838-1856},
          doi = {10.1093/mnras/stae2595},
archivePrefix = {arXiv},
       eprint = {2410.02875},
 primaryClass = {astro-ph.HE},
       adsurl = {https://ui.adsabs.harvard.edu/abs/2025MNRAS.536.1838B}
}

@ARTICLE{2015MNRAS.446.1939F,
       author = {{Feldmann}, Robert and {Mayer}, Lucio},
        title = "{The Argo simulation - I. Quenching of massive galaxies at high redshift as a result of cosmological starvation}",
      journal = {\mnras},
         year = 2015,
        month = jan,
       volume = {446},
       number = {2},
        pages = {1939-1956},
          doi = {10.1093/mnras/stu2207},
archivePrefix = {arXiv},
       eprint = {1404.3212},
 primaryClass = {astro-ph.GA},
       adsurl = {https://ui.adsabs.harvard.edu/abs/2015MNRAS.446.1939F}
}

@ARTICLE{2015MNRAS.446.1957F,
       author = {{Fiacconi}, Davide and {Feldmann}, Robert and {Mayer}, Lucio},
        title = "{The Argo simulation - II. The early build-up of the Hubble sequence}",
      journal = {\mnras},
         year = 2015,
        month = jan,
       volume = {446},
       number = {2},
        pages = {1957-1972},
          doi = {10.1093/mnras/stu2228},
archivePrefix = {arXiv},
       eprint = {1410.6818},
 primaryClass = {astro-ph.GA},
       adsurl = {https://ui.adsabs.harvard.edu/abs/2015MNRAS.446.1957F}
}

@ARTICLE{2012LRR....15....7C,
       author = {{Chru{\'s}ciel}, Piotr T. and {Costa}, Jo{\~a}o Lopes and {Heusler}, Markus},
        title = "{Stationary Black Holes: Uniqueness and Beyond}",
      journal = {Living Reviews in Relativity},
         year = 2012,
        month = dec,
       volume = {15},
       number = {1},
          eid = {7},
        pages = {7},
          doi = {10.12942/lrr-2012-7},
archivePrefix = {arXiv},
       eprint = {1205.6112},
 primaryClass = {gr-qc},
       adsurl = {https://ui.adsabs.harvard.edu/abs/2012LRR....15....7C}
}

@ARTICLE{2026A&A...713A...5S,
       author = {{Saha}, Navonil and {Sobolenko}, Margarita and {Berczik}, Peter and {Just}, Andreas and {Khan}, Fazeel Mahmood},
        title = "{The effect of triaxial galaxy shapes on the dynamics of triple supermassive black holes in a cosmological context}",
      journal = {\aap},
         year = 2026,
        month = aug,
       volume = {713},
          eid = {A5},
        pages = {A5},
          doi = {10.1051/0004-6361/202659844},
archivePrefix = {arXiv},
       eprint = {2607.23919},
 primaryClass = {astro-ph.GA},
       adsurl = {https://ui.adsabs.harvard.edu/abs/2026A&A...713A...5S}
}

@ARTICLE{2013MNRAS.434.2999B,
       author = {{Brem}, Patrick and {Amaro-Seoane}, Pau and {Spurzem}, Rainer},
        title = "{Relativistic mergers of compact binaries in clusters: the fingerprint of the spin}",
      journal = {\mnras},
         year = 2013,
        month = oct,
       volume = {434},
       number = {4},
        pages = {2999-3007},
          doi = {10.1093/mnras/stt1220},
archivePrefix = {arXiv},
       eprint = {1302.3135},
 primaryClass = {astro-ph.CO},
       adsurl = {https://ui.adsabs.harvard.edu/abs/2013MNRAS.434.2999B}
}

@ARTICLE{2004PhRvD..70j4011M,
       author = {{Memmesheimer}, Raoul-Martin and {Gopakumar}, Achamveedu and {Sch{\"a}fer}, Gerhard},
        title = "{Third post-Newtonian accurate generalized quasi-Keplerian parametrization for compact binaries in eccentric orbits}",
      journal = {\prd},
         year = 2004,
        month = nov,
       volume = {70},
       number = {10},
          eid = {104011},
        pages = {104011},
          doi = {10.1103/PhysRevD.70.104011},
archivePrefix = {arXiv},
       eprint = {gr-qc/0407049},
 primaryClass = {gr-qc},
       adsurl = {https://ui.adsabs.harvard.edu/abs/2004PhRvD..70j4011M}
}

@ARTICLE{Blanchet2024,
       author = {{Blanchet}, Luc},
        title = "{Post-Newtonian theory for gravitational waves}",
      journal = {Living Reviews in Relativity},
         year = 2024,
        month = dec,
       volume = {27},
       number = {1},
          eid = {4},
        pages = {4},
          doi = {10.1007/s41114-024-00050-z},
       adsurl = {https://ui.adsabs.harvard.edu/abs/2024LRR....27....4B}
}

@BOOK{Binney2008,
       author = {{Binney}, James and {Tremaine}, Scott},
        title = "{Galactic Dynamics: Second Edition}",
         year = 2008,
    publisher = {Princeton University Press, Princeton, NJ},
       adsurl = {https://ui.adsabs.harvard.edu/abs/2008gady.book.....B}
}

@ARTICLE{2013ApJ...764..184M,
       author = {{McConnell}, Nicholas J. and {Ma}, Chung-Pei},
        title = "{Revisiting the Scaling Relations of Black Hole Masses and Host Galaxy Properties}",
      journal = {\apj},
         year = 2013,
        month = feb,
       volume = {764},
       number = {2},
          eid = {184},
        pages = {184},
          doi = {10.1088/0004-637X/764/2/184},
archivePrefix = {arXiv},
       eprint = {1211.2816},
 primaryClass = {astro-ph.CO},
       adsurl = {https://ui.adsabs.harvard.edu/abs/2013ApJ...764..184M}
}

@ARTICLE{2017MNRAS.470.1121T,
       author = {{Tremmel}, M. and {Karcher}, M. and {Governato}, F. and {Volonteri}, M. and {Quinn}, T.~R. and {Pontzen}, A. and {Anderson}, L. and {Bellovary}, J.},
        title = "{The Romulus cosmological simulations: a physical approach to the formation, dynamics and accretion models of SMBHs}",
      journal = {\mnras},
         year = 2017,
        month = sep,
       volume = {470},
       number = {1},
        pages = {1121-1139},
          doi = {10.1093/mnras/stx1160},
archivePrefix = {arXiv},
       eprint = {1607.02151},
 primaryClass = {astro-ph.GA},
       adsurl = {https://ui.adsabs.harvard.edu/abs/2017MNRAS.470.1121T}
}

@ARTICLE{2005LRR.....8....8M,
       author = {{Merritt}, David and {Milosavljevi{\'c}}, Milos},
        title = "{Massive Black Hole Binary Evolution}",
      journal = {Living Reviews in Relativity},
         year = 2005,
        month = nov,
       volume = {8},
        pages = {8},
          doi = {10.12942/lrr-2005-8},
archivePrefix = {arXiv},
       eprint = {astro-ph/0410364},
 primaryClass = {astro-ph},
       adsurl = {https://ui.adsabs.harvard.edu/abs/2005LRR.....8....8M}
}

@ARTICLE{2018MNRAS.477.2310B,
       author = {{Bortolas}, E. and {Gualandris}, A. and {Dotti}, M. and {Read}, J.~I.},
        title = "{The influence of massive black hole binaries on the morphology of merger remnants}",
      journal = {\mnras},
         year = 2018,
        month = jun,
       volume = {477},
       number = {2},
        pages = {2310-2325},
          doi = {10.1093/mnras/sty775},
archivePrefix = {arXiv},
       eprint = {1710.04658},
 primaryClass = {astro-ph.GA},
       adsurl = {https://ui.adsabs.harvard.edu/abs/2018MNRAS.477.2310B}
}

@ARTICLE{1972ApJ...178..347B,
       author = {{Bardeen}, James M. and {Press}, William H. and {Teukolsky}, Saul A.},
        title = "{Rotating Black Holes: Locally Nonrotating Frames, Energy Extraction, and Scalar Synchrotron Radiation}",
      journal = {\apj},
         year = 1972,
        month = dec,
       volume = {178},
        pages = {347-370},
          doi = {10.1086/151796},
       adsurl = {https://ui.adsabs.harvard.edu/abs/1972ApJ...178..347B}
}

@ARTICLE{2005ApJ...620...69V,
       author = {{Volonteri}, Marta and {Madau}, Piero and {Quataert}, Eliot and {Rees}, Martin J.},
        title = "{The Distribution and Cosmic Evolution of Massive Black Hole Spins}",
      journal = {\apj},
         year = 2005,
        month = feb,
       volume = {620},
       number = {1},
        pages = {69-77},
          doi = {10.1086/426858},
archivePrefix = {arXiv},
       eprint = {astro-ph/0410342},
 primaryClass = {astro-ph},
       adsurl = {https://ui.adsabs.harvard.edu/abs/2005ApJ...620...69V}
}

@ARTICLE{2008ApJ...684..822B,
       author = {{Berti}, Emanuele and {Volonteri}, Marta},
        title = "{Cosmological Black Hole Spin Evolution by Mergers and Accretion}",
      journal = {\apj},
         year = 2008,
        month = sep,
       volume = {684},
       number = {2},
        pages = {822-828},
          doi = {10.1086/590379},
archivePrefix = {arXiv},
       eprint = {0802.0025},
 primaryClass = {astro-ph},
       adsurl = {https://ui.adsabs.harvard.edu/abs/2008ApJ...684..822B}
}

@BOOK{2003gnbs.book.....A,
       author = {{Aarseth}, Sverre J.},
        title = "{Gravitational N-Body Simulations}",
         year = 2003,
         publisher = {Cambridge University Press},
       adsurl = {https://ui.adsabs.harvard.edu/abs/2003gnbs.book.....A}
}
%%%%%%%%%%%%%%%%%%%%%%%%%%%%%%%%%%%%%%%%%%%%%%%%%%%%%%%%%%%%%%%%%%%%

%%%%%%%%%%%%%%%%%%%%%%%%%%%%%%%%%%%%%%%%%%%%%%%%%%%%%%%%%%%%%%%%%%%%
\begin{appendix}
\section{Statistics of stellar encounters}
\label{app:encounters}
This appendix gives supporting statistics for the stellar encounters analyzed in Sect.~\ref{Energy Balance of the binaries}.
An encounter is a passage of a star through a sphere of radius 10\,$a$ around the binary, with entry and exit times $t_{\rm in}$ and $t_{\rm out}$ as in Eq.~(\ref{Etot}), in the same spirit as the analysis of \citet{2021A&A...649A..41A}.
The energy change $\Delta E_*/m_*$ of a star during an encounter is defined in Sect.~\ref{Energy Balance of the binaries}.
The inclination $i$ of a star is derived from the orientation of its angular momentum vector relative to that of the binary, so that $\cos i \ge 0$ corresponds to prograde and $\cos i < 0$ to retrograde orbits.
\begin{figure}[h!]
    \centering
    \includegraphics[width=\columnwidth]{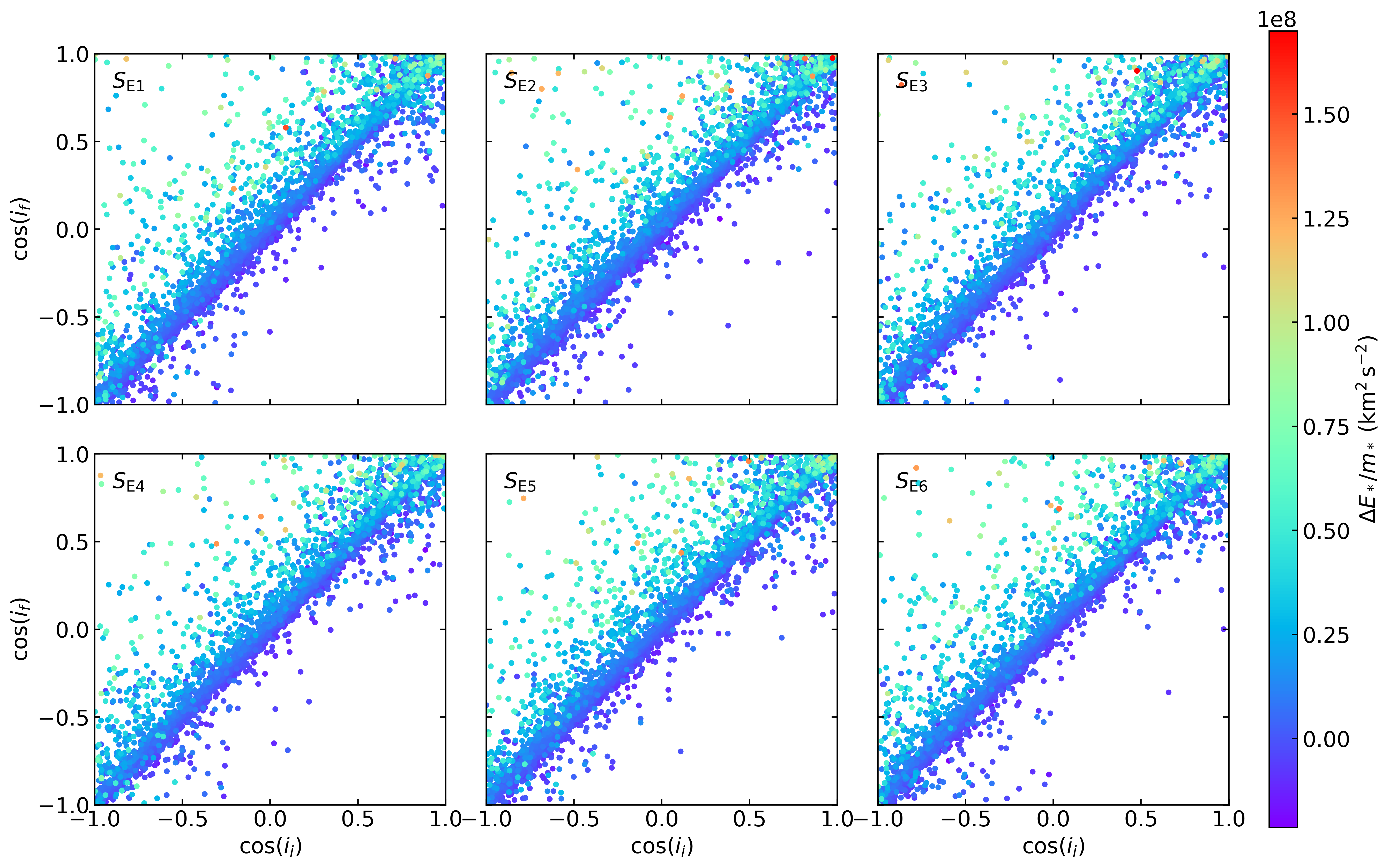}
    \caption{Same as bottom of Fig.~\ref{star-balance} but for all the other spinning models.}
    \label{fig:All_models_cosi_cosf}
\end{figure}

%%%%%%%%%%%%%%%%%%%%%%%%%%%%%%%%%%%%%%%%%%%%%%%%%%%%%%%%%%%%%%%%%%%%

%%%%%%%%%%%%%%%%%%%%%%%%%%%%%%%%%%%%%%%%%%%%%%%%%%%%%%%%%%%%%%%%%%%
\subsection{Inclination changes in the spinning models}
\label{app:inclination}

Fig.~\ref{app:inclination} repeats the bottom panel of Fig.~\ref{star-balance} for the spinning models $S_{\rm E1}$--$S_{\rm E6}$, while the non-spinning model $S_0$ is shown in Fig.~\ref{star-balance}.
Each panel shows the distribution of the initial against the final inclination of the interacting stars, color-coded by $\Delta E_*/m_*$.
In all models most stars lie along the diagonal, i.e. their orbital orientation is hardly changed by the encounter.
A smaller population lies above the diagonal at $\cos i_{\rm i} < 0$.
These are stars that arrive on retrograde orbits and leave on prograde ones, with intermediate energy gains.
This sign change of the angular momentum is also seen in the non-spinning model, in the same merger remnant at an earlier stage of the evolution \citep{2021A&A...649A..41A}, and in rotating nuclei \citep{2014ApJ...780..164W}.
It is consistent with the energy change of a star being proportional to the change of its angular-momentum component along the binary axis \citep{2017ApJ...837..135R}.
We find no qualitative difference between the spinning models and $S_0$, in line with the absence of a significant spin dependence in Sect.~\ref{sec:Results}.
Only a small number of encounters populate the high-energy region of each panel, so differences between individual panels there are not significant.

\subsection{Duration between the encounters}
\label{app:duration}

\begin{figure}[h!]
    \centering
    \includegraphics[width=\columnwidth]{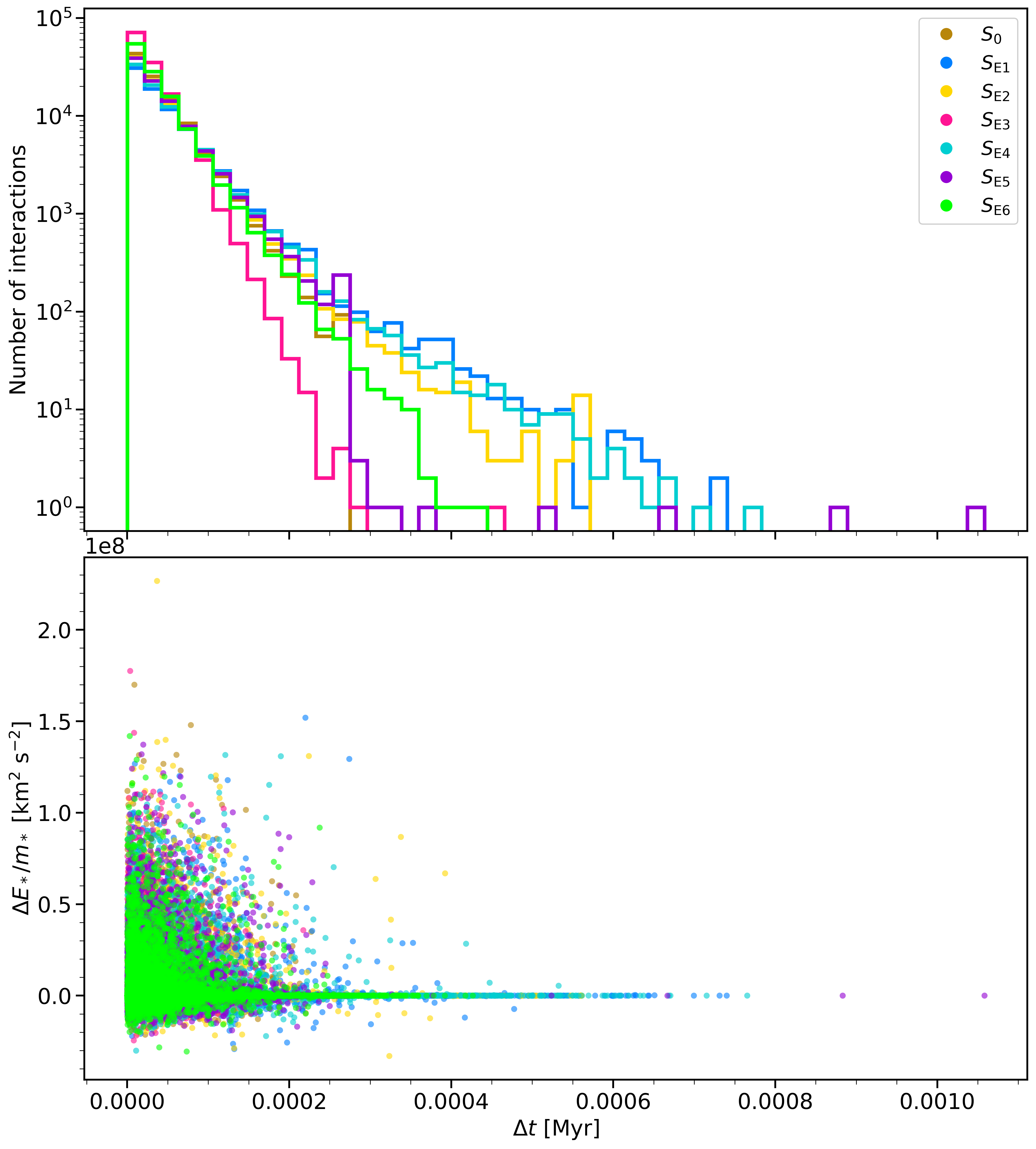}
    \caption{Duration between the stellar encounters. \textit{Top}: histogram of the number of encounters as a function of the time $\Delta t$. \textit{Bottom}: specific energy change $\Delta E_*/m_*$ of the later interaction as a function of $\Delta t$. Colors indicate the spin model.}
    \label{fig:N_dt}
\end{figure}

Fig.~\ref{fig:N_dt} shows the time $\Delta t$ between successive interactions in the time-ordered list of the stellar interactions of each run. The top panel gives the number of interactions as a function of $\Delta t$, and the bottom panel shows $\Delta E_*/m_*$ of the later interaction of each pair against $\Delta t$. The histogram falls steeply with increasing $\Delta t$, from of order $10^{5}$ per bin at the shortest waiting times to only a few per bin beyond $\Delta t \approx 5\cdot10^{-4}$\,Myr ($\approx 500$\,yr). The largest energy changes ($\Delta E_*/m_* \gtrsim 10^{8}\,\mathrm{km^{2}\,s^{-2}}$) occur only for waiting times $\Delta t \lesssim 2\cdot10^{-4}$\,Myr, whereas the long waiting times are associated with $\Delta E_*/m_* \approx 0$. The distributions are similar for all spin models. The tails beyond $\Delta t \approx 3\cdot10^{-4}$\,Myr contain only a handful of interactions per bin and may reflect differences in the total number of registered interactions between the runs, so we do not consider the differences there (e.g.\ between $S_{\rm E3}$ and $S_{\rm E5}$) significant.

\end{appendix}

\end{document}